\documentclass[journal]{IEEEtran}

\usepackage{changepage}
\usepackage{cite}
\usepackage{graphicx}
\usepackage{amsmath}  
\usepackage{algorithm,algpseudocode}
\usepackage{algorithmicx}
\usepackage{algpseudocode}
\usepackage{algorithm}
\usepackage{color}
\usepackage{verbatim}
\usepackage{amsfonts}
\usepackage{amsmath,amssymb}
\usepackage{url}
\usepackage{bbm}
\usepackage{bm}
\usepackage{subfig}
\usepackage{soul}

\DeclareMathOperator*{\maximize}{maximize}
\DeclareMathOperator{\subjectto}{subject~to}

\makeatletter
\def\endthebibliography{%
	\def\@noitemerr{\@latex@warning{Empty `thebibliography' environment}}%
	\endlist
}
\makeatother


\begin{document}
	\title{Multi-Agent Deep Reinforcement Learning for Dynamic Power Allocation in Wireless Networks}
	\author{Yasar Sinan Nasir, {\em Student Member, IEEE,} and Dongning Guo, {\em Senior Member, IEEE}\\
		\thanks{The authors are with Department of Electrical Engineering and Computer Science Northwestern University, Evanston, IL 60208. (e-mails: snasir@u.northwestern.edu; dguo@northwestern.edu)}}
	\maketitle
	\begin{abstract}
		This work demonstrates the potential of deep reinforcement learning techniques for transmit power control in wireless networks. Existing techniques typically find near-optimal power allocations by solving a challenging optimization problem. Most of these algorithms are not scalable to large networks in real-world scenarios because of their computational complexity and instantaneous cross-cell channel state information (CSI) requirement. In this paper, a distributively executed dynamic power allocation scheme is developed based on model-free deep reinforcement learning. Each transmitter collects CSI and quality of service (QoS) information from several neighbors and adapts its own transmit power accordingly. The objective is to maximize a weighted sum-rate utility function, which can be particularized to achieve maximum sum-rate or proportionally fair scheduling. Both random variations and delays in the CSI are inherently addressed using deep Q-learning. For a typical network architecture, the proposed algorithm is shown to achieve near-optimal power allocation in real time based on delayed CSI measurements available to the agents. 
		The proposed scheme is especially suitable for practical scenarios where the system model is inaccurate and CSI delay is non-negligible.
	\end{abstract}
	\section{Introduction}
	\label{sec:Intro}
	In emerging and future wireless networks, inter-cell interference management is one of the key technological challenges as access points (APs) become denser to meet ever-increasing demand on the capacity. A transmitter may increase its transmit power to improve its own data rate, but at the same time it may degrade links it interferes with. Transmit power control has been implemented since the first generation cellular networks \cite{chiang2007power}. Our goal here is to maximize an arbitrary weighted sum-rate objective, which achieves maximum sum-rate or proportionally fair scheduling as special cases.
	
	A number of centralized and distributed optimization techniques have been used to develop algorithms for reaching a suboptimal power allocation \cite{shi2011wmmse,shen2018fractional,illsoo2015distributedpower,chiang2007power,huang2006distributedpower,kiani2007distributed,zhang2011proportionally}. We select two state-of-the-art algorithms as benchmarks. These are the weighted minimum mean square error (WMMSE) algorithm \cite{shi2011wmmse} and an iterative algorithm based on fractional programming (FP) \cite{shen2018fractional}. In their generic form, both algorithms require full up-to-date cross-cell channel state information (CSI). To the best of our knowledge, this work is the first to apply deep reinforcement learning to power control \cite{nasir2018deep}. Sun \emph{et al.} \cite{sun2017learning} proposed a centralized {\em supervised learning} approach to train a fast deep neural network (DNN) that achieves 90\% or higher of the sum-rate achieved by the WMMSE algorithm. However, this approach still requires acquiring the full CSI. Another issue is that training DNN depends on a massive dataset of the WMMSE algorithm's output for randomly generated CSI matrices. Such a dataset takes a significant amount of time to produce due to WMMSE's computational complexity. As the network gets larger, the total number of DNN's input and output ports also increases, which raises questions on the scalability of the centralized solution of \cite{sun2017learning}. Furthermore, the success of supervised learning is highly dependent on the accuracy of the system model underlying the computed training data, which requires a new set of training data every time the system model or key parameters change.
	
	In this work, we design a distributively executed algorithm to be employed by all transmitters to compute their best power allocation in real time. Such a dynamic power allocation problem with time-varying channel conditions for a different system model and network setup was studied in \cite{neely2005dynamicpowercontrol} and the delay performance of the classical dynamic backpressure algorithm was improved by exploiting the stochastic Lyapunov optimization framework.
	
	The main contributions in this paper and some advantages of the proposed scheme are summarized as follows.
	\begin{enumerate}
		\item The proposed algorithm is one of the first power allocation schemes to use deep reinforcement learning in the literature. In particular, the distributively executed algorithm is based on deep Q-learning \cite{mnih2015human}, which is model-free and robust to unpredictable changes in the wireless environment.
		\item The complexity of the distributively executed algorithm does not depend on the network size. In particular, the proposed algorithm is computationally scalable to networks that cover arbitrarily large geographical areas if the number of links per unit area remains upper bounded by the same constant everywhere.
		\item The proposed algorithm learns a policy that guides all links to adjust their power levels under important practical constraints such as delayed information exchange and incomplete cross-link CSI.
		\item Unlike the supervised learning approach \cite{sun2017learning}, there is no need to run an existing near-optimal algorithm to produce a large amount of training data. We use an applicable centralized network trainer approach that gathers local observations from all network agents. This approach is computationally efficient and robust. In fact, a pretrained neural network can also achieve comparable performance as that of the centralized optimization based algorithms.
		\item We compare the reinforcement learning outcomes with state-of-the-art optimization-based algorithms. We also show the scalability and the robustness of the proposed algorithm using simulations. In the simulation, we model the channel variations inconsequential to the learning algorithm using the Jakes fading model \cite{liang2017delayedCSI}. In certain scenarios the proposed distributed algorithm even outperforms the centralized iterative algorithms introduced in \cite{shi2011wmmse, shen2018fractional}. We also address some important practical constraints that are not included in \cite{shi2011wmmse, shen2018fractional}.
	\end{enumerate}
	
	Deep reinforcement learning framework has been used in some other wireless communications problems \cite{luong2018deepRLsurvey,ye2017deep,yu2017deep,zhao2018slicing}. Classical Q-learning techniques have been applied to the power allocation problem in \cite{bennis2010qlearning,simsek2011qtable,amiri2018mlpower,ghadimi2017dynamicpower,calabrese2017learning}. The goal in \cite{bennis2010qlearning, simsek2011qtable} is to reduce the interference in LTE-Femtocells. Unlike the \emph{deep} Q-learning algorithm, the classical algorithm builds a lookup table to represent the value of state-action pairs, so \cite{bennis2010qlearning} and \cite{simsek2011qtable} represent the wireless environment using a discrete state set and limit the number of learning agents. Amiri \emph{et al.} \cite{amiri2018mlpower} have used cooperative Q-learning based power control to increase the QoS of users in femtocells without considering the channel variations. The deep Q-learning based power allocation to maximize the network objective has also been considered in \cite{ghadimi2017dynamicpower,calabrese2017learning}. Similar to the proposed approach, the work in \cite{ghadimi2017dynamicpower,calabrese2017learning} is also based on a distributed framework with a centralized training assumption, but the benchmark to evaluate the performance of their algorithm was a fixed power allocation scheme instead of state-of-the-art algorithms. The proposed approach to the state of wireless environment and the reward function is also novel and unique. Specifically, the proposed approach addresses the stochastic nature of wireless environment as well as incomplete/delayed CSI, and arrives at highly competitive strategies quickly. 
	
	The remainder of this paper is organized as follows. We give the system model in Section \ref{sec:model}. In Section \ref{sec:dynpowcontrol}, we formulate the dynamic power allocation problem and give our practical constraints on the local information. In Section \ref{sec:DQN}, we first give an overview of deep Q-learning and then describe the proposed algorithm. We give simulation results in Section \ref{sec:resutls}. We conclude with a discussion of possible future work in Section \ref{sec:conclusion}.	
	\section{System Model}\label{sec:model}
	\label{s:math}	
	We first consider the classical power allocation problem in a network of $n$ links. We assume that all transmitters and receivers are equipped with a single antenna. The model is often used to describe a mobile ad hoc network (MANET) \cite{huang2006distributedpower}. The model has also been used to describe a simple cellular network with $n$ APs, where each AP serves a single user device \cite{shen2018fractional,illsoo2015distributedpower}. Let $N = \left\{1,\dots,n\right\}$ denote the set of link indexes. We consider a fully synchronized time slotted system with slot duration $T$. For simplicity, we consider a single frequency band with flat fading. We adopt a block fading model to denote the downlink channel gain from transmitter $i$ to receiver $j$ in time slot $t$ as
	\begin{align}\label{eq:channel}
	g^{(t)}_{i\rightarrow j} &= \left| h^{(t)}_{i\rightarrow j}\right|^2 \alpha_{i\rightarrow j}, \quad t=1,2,\dots .
	\end{align}
	Here, $\alpha_{i\rightarrow j} \geq 0$ represents the large-scale fading component including path loss and log-normal shadowing, which remains the same over many time slots. Following Jakes fading model \cite{liang2017delayedCSI}, we express the small-scale Rayleigh fading component as a first-order complex Gauss-Markov process:
	\begin{align}\label{eq:JakesModel}
	h_{i\rightarrow j}^{(t)} &= \rho h_{i\rightarrow j}^{(t-1)} + \sqrt{1-\rho^2}e_{i\rightarrow j}^{(t)}
	\end{align}
	where $h_{i\rightarrow j}^{(0)}$ and the channel innovation process $e_{i\rightarrow j}^{(1)},e_{i\rightarrow j}^{(2)},\dots$ are independent and identically distributed circularly symmetric complex Gaussian (CSCG) random variables with unit variance. The correlation $\rho=J_0(2\pi f_d T)$, where $J_0(.)$ is the zeroth-order Bessel function of the first kind and $f_d$ is the maximum Doppler frequency. 
	
	The received signal-to-interference-plus-noise ratio (SINR) of link $i$ in time slot $t$ is a function of the allocation $\bm{p}=\left[p_1,\dots,p_n\right]^\intercal$:
	\begin{align}\label{eq:SINR}
	\gamma^{(t)}_{i}\left(\bm{p}\right) &= \frac{g^{(t)}_{i\rightarrow i}p_i}{\sum_{j \neq i}g^{(t)}_{j\rightarrow i}p_j+\sigma^2}
	\end{align}
	where $\sigma^2$ is the additive white Gaussian noise (AWGN) power spectral density (PSD). We assume the same noise PSD in all receivers without loss of generality. The downlink spectral efficiency of link $i$ at time $t$ can be expressed as:
	\begin{align}\label{eq:DynRate}
	\begin{split}
	C^{(t)}_i \left(\bm{p}\right) &= \log\left(1+\gamma^{(t)}_{i}\left(\bm{p}\right)\right).
	\end{split}
	\end{align}
	
	The transmit power of transmitter $i$ in time slot $t$ is denoted as $p^{(t)}_i$. We denote the power allocation of the network in time slot $t$ as $\bm{p}^{(t)}=\left[p^{(t)}_1,\dots,p^{(t)}_n\right]^\intercal$.
	\section{Dynamic Power Control}\label{sec:dynpowcontrol}\label{sec:practical}
	We are interested in maximizing a generic weighted sum-rate objective function. Specifically, the dynamic power allocation problem in slot $t$ is formulated as
	\begin{align}\label{eq:DynOptProblem}
	\begin{split}
	\maximize_{\bm{p}} & \quad \sum_{i=1}^{n}w_i^{(t)} \cdot C^{(t)}_i \left(\bm{p}\right) \\
	\subjectto & \quad 0 \leq p_i \leq P_{\textrm{max}} ,\quad i = 1, \dots, n \,,
	\end{split}
	\end{align}
	where $w_i^{(t)}$ is the given nonnegative weight of link $i$ in time slot $t$, and $P_{\textrm{max}}$ is the maximum PSD a transmitter can emit. Hence, the dynamic power allocator has to solve an independent problem in the form of \eqref{eq:DynOptProblem} at the beginning of every time slot. In time slot $t$, the optimal power allocation solution is denoted as $\bm{p}^{(t)}$. Problem \eqref{eq:DynOptProblem} is in general non-convex and has been shown to be NP-hard \cite{Luo2008dynamicspectrum}.
	
	We consider two special cases. In the first case, the objective is to maximize the sum-rate by letting $w_i^{(t)}=1$ for all $i$ and $t$. In the second case, the weights vary in a controlled manner to ensure proportional fairness \cite{tse2005fundamentals,zhang2011proportionally}. Specifically, at the end of time slot $t$, receiver $i$ computes its weighted average spectral efficiency as
	\begin{align}\label{eq:exponentialmovingaveragerate}
	\begin{split}
	\bar{C}^{(t)}_i &= \beta \cdot C^{(t)}_i \left(\bm{p}^{(t)}\right) + (1-\beta)\bar{C}^{(t-1)}_i
	\end{split}
	\end{align}
	where $\beta \in (0,1]$ is used to control the impact of history. User $i$ updates its link weight as:
	\begin{align}\label{eq:pfsweight}
	\begin{split}
	w_i^{(t+1)} &= \left(\bar{C}^{(t)}_i\right)^{-1} .
	\end{split}
	\end{align}
	This power allocation algorithm maximizes the sum of log-average spectral efficiency \cite{tse2005fundamentals}, i.e.,
	\begin{align}\label{eq:pfsobjective}
	\begin{split}
	\sum_{i\in N}\log{\bar{C}^{(t)}_i}
	\end{split},
	\end{align}
	where a user's long-term average throughput is proportional to its long-term channel quality in some sense.
	
	We use two popular (suboptimal) power allocation algorithms as benchmarks. These are the WMMSE algorithm \cite{shi2011wmmse} and the FP algorithm \cite{shen2018fractional}. Both are centralized and iterative in their original form. The closed-form FP algorithm used in this paper is formulated in \cite[Algorithm 3]{shen2018fractional}. Similarly, a detailed explanation and pseudo code of the WMMSE algorithm is given in \cite[Algorithm 1]{sun2017learning}.
	The WMMSE and FP algorithms are both centralized and require full cross-link CSI. The centralized mechanism is suitable for a stationary environment with slowly varying weights and no fast fading. For a network with non-stationary environment, it is infeasible to instantaneously collect all CSI over a large network.
	
	It is fair to assume that the feedback delay $T_{\textrm{fb}}$ from a receiver to its corresponding transmitter is much smaller than the slot duration $T$, so the prediction error due to the feedback delay is neglected. Therefore, once receiver $i$ completes a direct channel measurement, we assume that it is also available at the transmitter $i$.
	
	For the centralized approach, once a link acquires the CSI of its direct channel and all other interfering channels to its receiver, passing this information to a central controller is another burden. This is typically resolved using a backhaul network between the APs and the central controller. The CSI of cross links is usually delayed or even outdated. Furthermore, the central controller can only return the optimal power allocation as the iterative algorithm converges, which is another limitation on the scalability.
	
	Our goal is to design a scalable algorithm, so we limit the information exchange to between nearby transmitters. We define two neighborhood sets for every $i\in N$: Let the set of transmitters whose SNR at receiver $i$ was above a certain threshold $\eta$ during the past time slot $t-1$ be denoted as
	\begin{align}\label{eq:InNeigh}
	I^{(t)}_{i} = \left\{j\in N, j \neq i \middle| g^{(t-1)}_{j\rightarrow i}p^{(t-1)}_j  > \eta \sigma^2 \right\}.
	\end{align}
	Let the set of receiver indexes whose SNR from transmitter $i$ was above a threshold in slot $t-1$ be denoted as
	\begin{align}\label{eq:OutNeigh}
	O^{(t)}_{i} = \left\{k \in N, k \neq i \middle| g^{(t-1)}_{i\rightarrow j}p^{(t-1)}_i  > \eta \sigma^2 \right\}.
	\end{align}
	From link $i$'s viewpoint, $I^{(t)}_{i}$ represents the set of ``interferers'', whereas $O^{(t)}_{i}$ represents the set of the ``interfered'' neighbors.
	
	We next discuss the local information a transmitter possesses at the beginning of time slot $t$. First, we assume that transmitter $i$ learns via receiver feedback the direct downlink channel gain, $g^{(t)}_{i\rightarrow i}$. Further, transmitter $i$ also learns the current total received interference-plus-noise power at receiver $i$ before the global power update, i.e., $\sum_{j \in N, j \neq i}g^{(t)}_{j\rightarrow i}p^{(t-1)}_j + \sigma^2$ (as a result of the new gains and the yet-to-be-updated powers). In addition, by the beginning of slot $t$, receiver $i$ has informed transmitter $i$ of the received power from every interferer $j \in I^{(t)}_{i}$, i.e., $g^{(t)}_{j\rightarrow i}p^{(t-1)}_j$. These measurements can only be available at transmitter $i$ just before the beginning of slot $t$. Hence, in the previous slot $t-1$, receiver $i$ also informs transmitter $i$ of the outdated versions of these measurements to be used in the information exchange process performed in slot $t-1$ between transmitter $i$ and its interferers. 
	\begin{figure}
		[t]
		\centering
		\includegraphics[clip, trim=0.25cm 0.25cm 3.75cm 0.25cm,width=1.0\columnwidth]{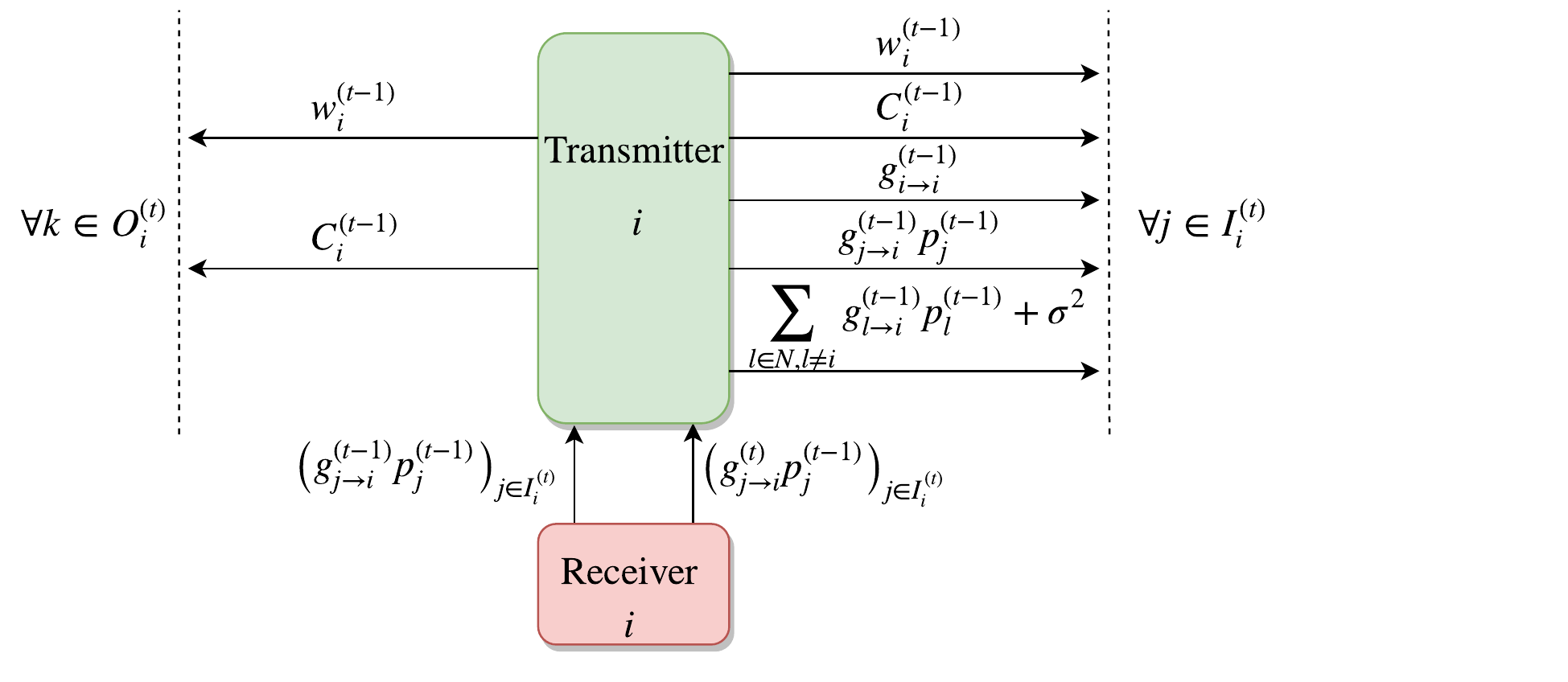}
		\caption{The information exchange between transmitter $i$ and its neighbors in time slot $t-1$. Note that transmitter $i$ obtains $g^{(t)}_{j\rightarrow i}p^{(t-1)}_j$ by the end of slot $t-1$, but it is not able to deliver this information to interferer $j$ before the beginning of slot $t$ due to additional delays through the backhaul network.}
		\label{fig:infoexchange}
	\end{figure}
	To clarify, as shown in Fig. \ref{fig:infoexchange}, transmitter $i$ has sent the following outdated information to interferer $j \in I^{(t)}_{i}$ in return for $w^{(t-1)}_j$ and $C^{(t-1)}_j$:
	\begin{itemize}
		\item the weight of link $i$, $w^{(t-1)}_i$,
		\item the spectral efficiency of link $i$ computed from \eqref{eq:DynRate}, $C^{(t-1)}_i$,
		\item the direct gain, $g^{(t-1)}_{i\rightarrow i}$,
		\item the received interference power from transmitter $j$, $g^{(t-1)}_{j\rightarrow i}p^{(t-1)}_j$,
		\item the total interference-plus-noise power at receiver $i$, i.e., $\sum_{l \in N, l \neq i}g^{(t-1)}_{l\rightarrow i}p^{(t-1)}_l + \sigma^2$.
	\end{itemize}
	As assumed earlier, these measurements are accurate, where the uncertainty about the current CSI is entirely due to the latency of information exchange (one slot). By the same token, from every interfered $k \in O^{(t)}_{i}$, transmitter $i$ also obtains $k$'s items listed above.	
	
	\section{Deep Reinforcement Learning for Dynamic Power Allocation}\label{sec:DQN}
	\subsection{Overview of Deep Q-Learning}\label{sec:singleagentDQN}
	A reinforcement learning agent learns its best policy from observing the rewards of trial-and-error interactions with its environment over time\cite{kaelbling1996reinforcement,sutton1998reinforcement}. Let $S$ denote a set of possible states and $A$ denote a discrete set of actions. The state $s \in S$ is a tuple of environment's features that are relevant to the problem at hand and it describes agent's relation with its environment \cite{ghadimi2017dynamicpower}. Assuming discrete time steps, the agent observes the state of its environment, $s^{(t)} \in S$ at time step $t$. It then takes an action $a^{(t)} \in A$ according to a certain policy $\pi$. The policy $\pi (s,a)$ is the probability of taking action $a$ conditioned on the current state being $s$. The policy function must satisfy $\sum_{a \in A}\pi (s,a) = 1$. Once the agent takes an action $a^{(t)}$, its environment moves from the current state $s^{(t)}$ to the next state $s^{(t+1)}$. As a result of this transition, the agent gets a reward $r^{(t+1)}$ that characterizes its benefit from taking action $a^{(t)}$ at state $s^{(t)}$. This scheme forms an experience at time $t+1$, hereby defined as $e^{(t+1)}=\left(s^{(t)},a^{(t)},r^{(t+1)},s^{(t+1)}\right)$, which describes an interaction with the environment \cite{mnih2015human}.
	
	The well-known Q-learning algorithm aims to compute an optimal policy $\pi$ that maximizes a certain expected reward without knowledge of the function form of the reward and the state transitions. Here we let the reward be the future cumulative discounted reward at time $t$:
	\begin{align}\label{eq:discountedreward}
	\begin{split}
	R^{(t)} &= \sum_{\tau=0}^{\infty}\gamma^{\tau} r^{(t+\tau+1)}
	\end{split}
	\end{align}
	where $\gamma \in (0,1]$ is the discount factor for future rewards. In the stationary setting, we define a Q-function associated with a certain policy $\pi$ as the expected reward once action $a$ is taken under state $s$\cite{singh2000convergence}, i.e.,
	\begin{align}\label{eq:qfunction}
	Q^{\pi}(s,a) &= \mathbb{E}_{\pi}\left[R^{(t)} \middle| s^{(t)}=s, a^{(t)}=a \right].
	\end{align} 
	As an action value function, the Q-function satisfies a Bellman equation \cite{serrano2010qlearning}:
	\begin{align}\label{eq:bellmaneq}
	Q^\pi(s,a)=\mathcal{R}(s,a)+\gamma \sum_{s' \in S}\mathcal{P}^{a}_{ss'}\left(\sum_{a' \in A}\pi(s',a')Q^\pi\left(s',a'\right)\right)
	\end{align}
	where $\mathcal{R}(s,a)=\mathbb{E} \left[r^{(t+1)} \middle|s^{(t)}=s,a^{(t)}=a \right]$ is the expected reward of taking action $a$ at state $s$, and $\mathcal{P}^{a}_{ss'} = \Pr\left(s^{(t+1)}=s'\middle| s^{(t)} =s, a^{(t)}=a\right)$ is the transition probability from given state $s$ to state $s'$ with action $a$. From the fixed-point equation \eqref{eq:bellmaneq}, the value of $(s,a)$ can be recovered from all values of $(s',a') \in S\times A$. It has been proved that some iterative approaches such as Q-learning algorithm efficiently converges to the action value function \eqref{eq:qfunction} \cite{singh2000convergence}. Clearly, it suffices to let $\pi^*(s,a)$ be equal to 1 for the most favorable action. From \eqref{eq:bellmaneq}, the optimal Q-function associated with the optimal policy is then expressed as
	\begin{align}\label{eq:bellmaneqopt}
	Q^*(s,a)=\mathcal{R}(s,a)+\gamma \sum_{s' \in S}\mathcal{P}^{a}_{ss'}\max_{a'}Q^*(s',a').
	\end{align}
	
	The classical Q-learning algorithm constructs a lookup table, $q(s,a)$, as a surrogate of the optimal Q-function. Once this lookup table is randomly initialized, the agent takes actions according to the $\epsilon$-greedy policy for each time step. The $\epsilon$-greedy policy implies that with probability $1-\epsilon$ the agent takes the action $a^*$ that gives the maximum lookup table value for a given current state, whereas it picks a random action with probability $\epsilon$ to avoid getting stuck at non-optimal policies \cite{mnih2015human}. After acquiring a new experience as a result of the taken action, the Q-learning algorithm updates a corresponding entry of the lookup table according to:
	\begin{align}\label{eq:qupdate}
	\begin{split}
	q\left(s^{(t)},a^{(t)}\right) &\gets (1-\alpha)q\left(s^{(t)},a^{(t)}\right) \\
	&\qquad + \alpha \left(r^{(t+1)} + \gamma \max_{a'}q\left(s^{(t+1)},a'\right)\right)
	\end{split}
	\end{align}
	where $\alpha \in (0,1]$ is the learning rate \cite{singh2000convergence}.
	
	In case the state and action spaces are very large, as is the case for the power control problem at hand. The classical Q-learning algorithm fails mainly because of two reasons:
	\begin{enumerate}\item Many states are rarely visited, and \item the storage of lookup table in \eqref{eq:qupdate} becomes impractical \cite{naparstek2017deep}. \end{enumerate} Both issues can be solved with deep reinforcement learning, e.g., deep Q-learning \cite{mnih2015human}. A deep neural network called deep Q-network (DQN) is used to estimate the Q-function in lieu of a lookup table. The DQN can be expressed as $q(s,a,\bm{\theta})$, where the real-valued vector $\bm{\theta}$ represents its parameters. The essence of DQN is that the function $q(\cdot,\cdot,\bm{\theta})$ is completely determined by $\bm{\theta}$. As such, the task of finding the best Q-function in a functional space of uncountably many dimensions is reduced to searching the best $\bm{\theta}$ of finite dimensions. Similar to the classical Q-learning, the agent collects experiences with its interaction with the environment. The agent or the network trainer forms a data set $D$ by collecting the experiences until time $t$ in the form of $(s,a,r',s')$. As the ``quasi-static target network'' method \cite{mnih2015human} implies, we define two DQNs: the target DQN with parameters $\bm{\theta}_{\textrm{target}}^{(t)}$ and the train DQN with parameters $\bm{\theta}_{\textrm{train}}^{(t)}$. $\bm{\theta}_{\textrm{target}}^{(t)}$ is updated to be equal to $\bm{\theta}_{\textrm{train}}^{(t)}$ once every $T_{u}$ steps. From the ``experience replay'' \cite{mnih2015human}, the least squares loss of train DQN for a random mini-batch $D^{(t)}$ at time $t$ is
	\begin{align}\label{eq:loss}
	L\left(\bm{\theta}_{\textrm{train}}^{(t)}\right) &= \sum_{(s,a,r',s') \in D^{(t)}} \left( y^{(t)}_{DQN}(r',s') -q\left(s,a;\bm{\theta}_{\textrm{train}}^{(t)}\right)\right)^2
	\end{align}
	where the target is 
	\begin{align}\label{eq:target}
	y^{(t)}_{DQN}(r',s') = r' + \lambda \max_{a'}q\left(s',a';\bm{\theta}_{\textrm{target}}^{(t)}\right).
	\end{align}
	Finally, we assume that each time step the stochastic gradient descent algorithm that minimizes the loss function \eqref{eq:loss} is used to train the mini-batch $D^{(t)}$. The stochastic gradient descent returns the new parameters of train DQN using the gradient computed from just few samples of the dataset and has been shown to converge to a set of good parameters quickly \cite{lecun2015deep}.		
	\subsection{Proposed Multi-Agent Deep Reinforcement Learning Algorithm}\label{sec:proposedalg}
\begin{figure}
		[t]
		\centering
		\includegraphics[clip, trim=5.25cm 0.0cm 6.5cm 0.0cm,width=1.0\columnwidth]{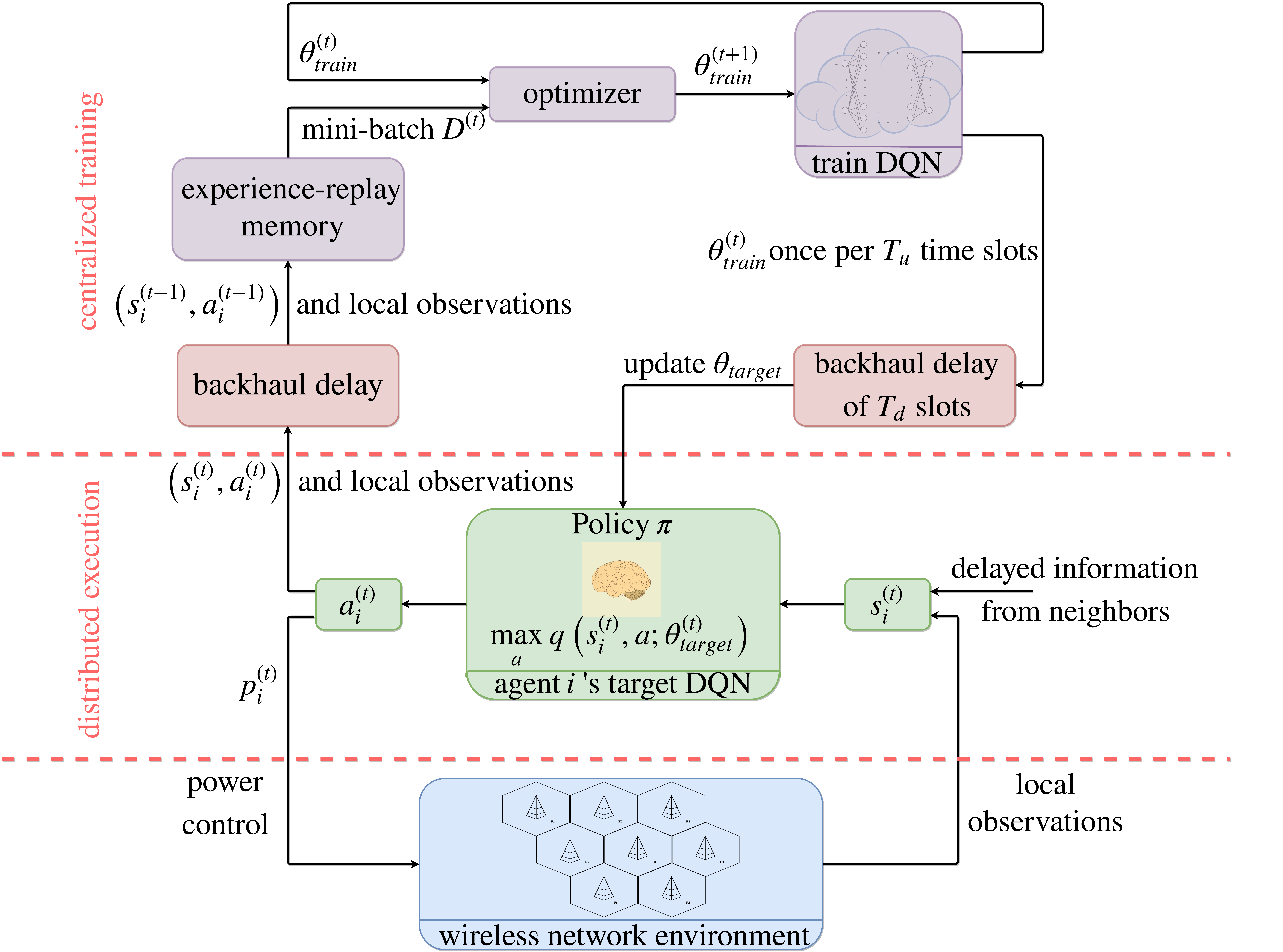}
		\caption{Illustration of the proposed multi-agent deep reinforcement learning algorithm.}
		\label{fig:systemmodel}
	\end{figure}

	As depicted in Fig. \ref{fig:systemmodel}, we propose a multi-agent deep reinforcement learning scheme with each transmitter as an agent. Similar to \cite{hu1998online}, we define the local state of learning agent $i$ as $s_i \in S_i$ which is composed of environment features that are relevant to agent $i$'s action $a_i \in A_i$. 
	In the multi-agent learning system, the state transitions of their common environment depend on the agents' joint actions. An agent's environment transition probabilities in \eqref{eq:bellmaneq} may not be stationary as other learning agents update their policies. The Markov property introduced for the single-agent case in Section \ref{sec:singleagentDQN} no longer holds in general \cite{nguyen2018multisurvey}. This ``environment non-stationarity'' issue may cause instability during the learning process. One way to tackle the issue is to train a single meta agent with a DQN that outputs joint actions for the agents \cite{foerster2017stabilising}. The complexity of the state-action space, and consequently the DQN complexity, will then be proportional to the total number of agents in the system. The single-meta agent approach is not suitable for our dynamic setup and the distributed execution framework, since its DQN can only forward the action assignments to the transmitters after acquiring the global state information. There is an extensive research to develop multi-agent learning frameworks and there exists several multi-agent Q-learning adaptations \cite{tampuu2017multiagent,nguyen2018multisurvey}. However, multi-agent learning is an open research area and theoretical guarantees for these adaptations are rare and incomplete despite their good empirical performances \cite{tampuu2017multiagent,nguyen2018multisurvey}. 
	
	In this work, we take an alternative approach where the DQNs are distributively executed at the transmitters, whereas training is centralized to ease implementation and to improve stability. Each agent $i$ has the same copy of the DQN with parameters $Q_{\textrm{target}}^{(t)}$ at time slot $t$. The centralized network trainer trains a single DQN by using the experiences gathered from all agents. This significantly reduces the amount of memory and computational resources required by training. The centralized training framework is also similar to the \emph{parameter sharing} concept which allows the learning algorithm to draw advantage from the fact that agents are learning together for faster convergence \cite{gupta2017cooperative}. Since agents are working collaboratively to maximize the global objective in \eqref{eq:DynOptProblem} with an appropriate reward function design to be discussed in Section \ref{sec:reward}, each agent can benefit from experiences of others. Note that sharing the same DQN parameters still allows different behavior among agents, because they execute the same DQN with different local states as input.
	
	As illustrated in Fig. \ref{fig:systemmodel}, at the beginning of time slot $t$, agent $i$ takes action $a_i^{(t)}$ as a function of $s_i^{(t)}$ based on the current decision policy. All agents are synchronized and take their actions at the same time. Prior to taking action, agent $i$ has observed the effect of the past actions of its neighbors on its current state, but it has no knowledge of $a^{(t)}_j$, $\forall j \neq i$. From the past experiences, agent $i$ is able to acquire an estimation of what is the impact of its own actions on future actions of its neighbors, and it can determine a policy that maximizes its discounted expected future reward with the help of deep Q-learning.
	
	\begin{figure}
		[t]
		\centering
		\subfloat[The illustration of all five layers of the proposed DQN: The input layer is followed by three hidden layers and an output layer. The notation $n$, $\omega$ and $b$ indicate DQN neurons, weights, and biases, respectively. These weights and biases form the set of DQN parameters denoted as $\theta$. The biases are not associated with any neuron and we multiply these biases by the scalar value 1.]{
			\includegraphics[clip, trim=0.5cm 0cm 2cm 0cm,width=1.0\columnwidth]{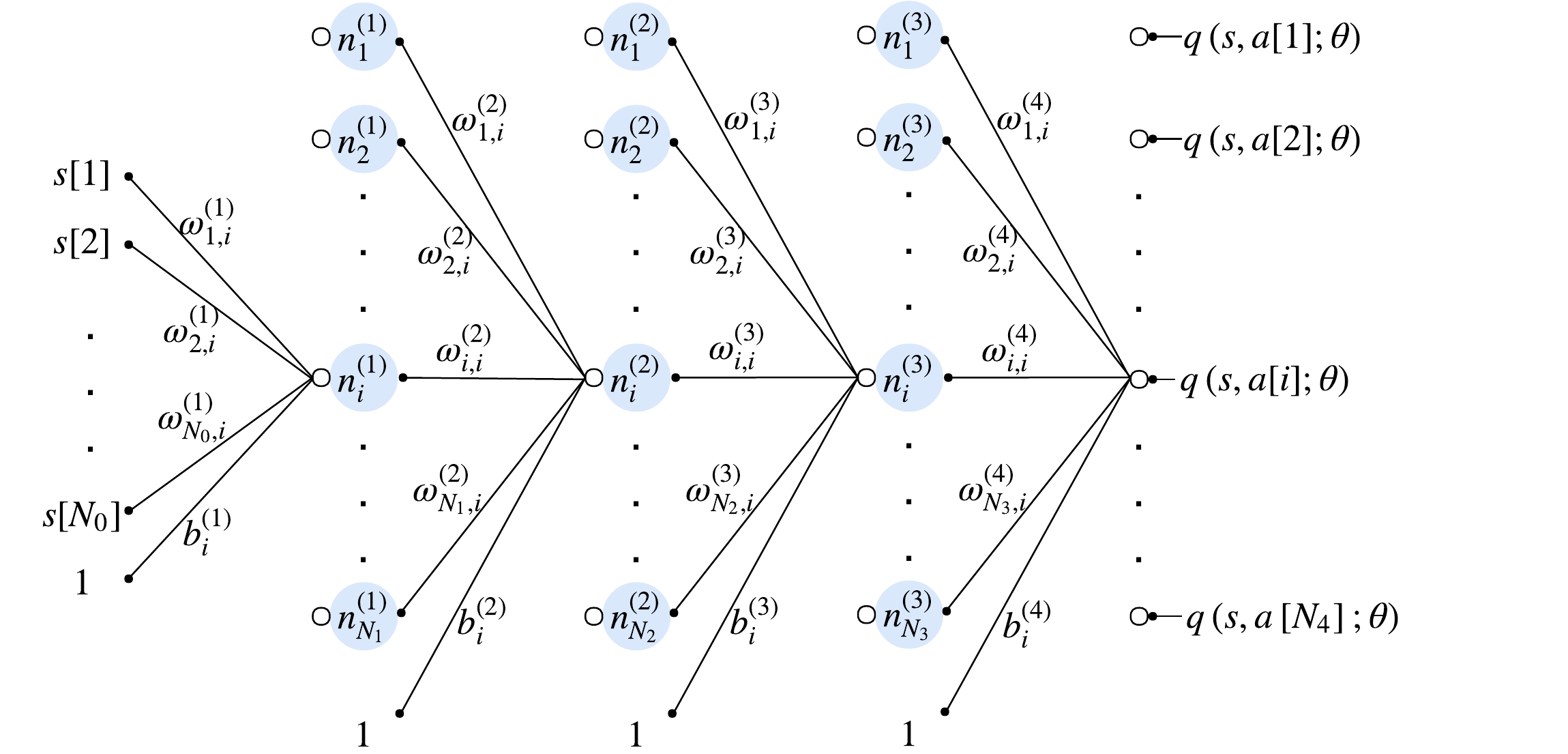}
			\label{fig:dqnmodela}}
		\hfil
		\subfloat[The functionality of a single neuron extracted from the first hidden-layer. $a(.)$ denotes the non-linear activation function. ]{
			\includegraphics[clip, trim=0.0cm 0.0cm 3.25cm 0.25cm,width=1.0\columnwidth]{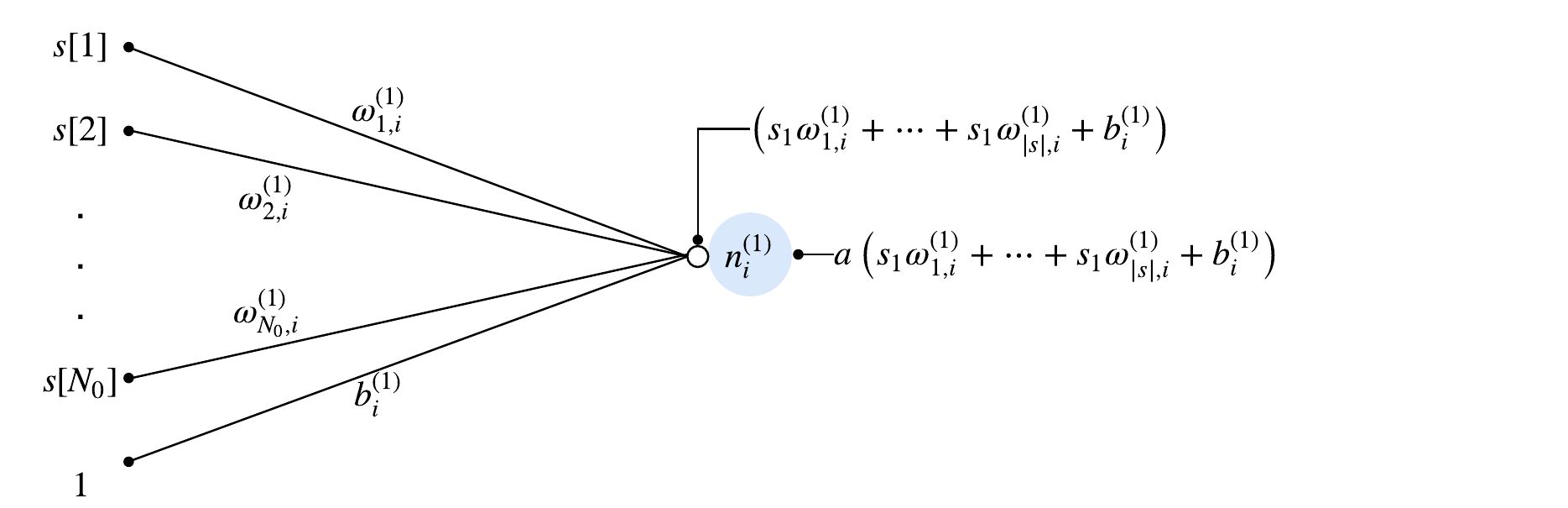}
			\label{fig:dqnmodelb}}
		\caption{The overall design of the proposed DQN.}
		\label{fig:dqnmodel}
	\end{figure}
	
	The proposed DQN is a fully-connected deep neural network \cite[Chapter~5]{watt2016machine} that consists of five layers as shown in Fig. \ref{fig:dqnmodela}. The first layer is fed by the input state vector of length $N_0$. We relegate the detailed design of the state vector elements to Section \ref{sec:states}. The input layer is followed by three hidden layers with $N_1$, $N_2$, and $N_3$ neurons, respectively. At the output layer, each port gives an estimate of the Q-function with given state input and the corresponding action output. The total number of DQN output ports is denoted as $N_4$ which is equal to the cardinality of the action set to be described in Section \ref{sec:actions}. The agent finds the action that has the maximum value at the DQN output and takes this action as its transmit power. 
	
	In Fig. \ref{fig:dqnmodela}, we also depicted the connection between these layers by using the weights and biases of the DQN which form the set of parameters. The total number of scalar parameters in the fully connected DQN is 
	\begin{align}\label{eq:DQNparameters}
	|\theta| = \sum_{l=0}^{3} \left(N_l + 1\right)N_{l+1}.
	\end{align}
	In addition, Fig. \ref{fig:dqnmodelb} describes the functionality of a single neuron which applies a non-linear activation function to its combinatorial input.
	
	During the training stage, in each time slot, the trainer randomly selects a mini-batch $D^{(t)}$ of $M_{b}$ experiences from an experience-replay memory \cite{mnih2015human} that stores the experiences of all agents. The experience-replay memory is a FIFO queue \cite{yu2017deep} with a length of $nM_{m}$ samples where $n$ is the total number of agents, i.e., a new experience replaces the oldest experience in the queue and the queue length is proportional to the number of agents. At time slot $t$ the most recent experience from agent $i$ is $e_i^{(t-1)}= \left(s_i^{(t-2)},a_i^{(t-2)},r_i^{(t-1)},s_i^{(t-1)}\right)$ due to delay. Once the trainer picks $D^{(t)}$, it updates the parameters to minimize the loss in \eqref{eq:loss} using an appropriate optimizer, e.g., the stochastic gradient descent method \cite{lecun2015deep}. As also explained in Fig. \ref{fig:systemmodel}, once per $T_{u}$ time slots, the trainer broadcasts the latest trained parameters. The new parameters are available at the agents after $T_{d}$ time slots due to the transmission delay through the backhaul network. Training may be terminated once the parameters converge.
	
	\subsection{States}\label{sec:states}
	As described in Section \ref{sec:practical}, agent $i$ builds its state $s_i^{(t)}$ using information from the interferer and interfered sets given by \eqref{eq:InNeigh} and \eqref{eq:OutNeigh}, respectively. To better control the complexity, we set $\left|\bar{I}^{(t)}_{i}\right|=\left|\bar{O}^{(t)}_{i}\right|=c$, where $c > 0$ is the restriction on the number of interferers and interfereds the AP communicating with. At the beginning of time slot $t$, agent $i$ sorts its interferers by current received power from interferer $j \in I^{(t)}_{i}$ at receiver $i$, i.e., $g^{(t)}_{j\rightarrow i}p^{(t-1)}_j$. This sorting process allows agent $i$ to prioritize its interferers. As $\left|I^{(t)}_{i}\right| > c$, we want to keep strong interferers which have higher impact on agent $i$'s next action. On the other hand, if $\left|I^{(t)}_{i}\right| < c$, agent $i$ adds $\left|I^{(t)}_{i}\right| - c$ virtual noise agents to $I^{(t)}_{i}$ to fit the fixed DQN. A virtual noise agent is assigned an arbitrary negative weight and spectral efficiency. Its downlink and interfering channel gains are taken as zero in order to avoid any impact on agent $i$'s decision-making. The purpose of having these virtual agents as placeholders is to provide inconsequential inputs to fill the input elements of fixed length, like `padding zeros'. After adding virtual noise agents (if needed), agent $i$ takes first $c$ interferers to form $\bar{I}^{(t)}_{i}$. For the interfered neighbors, agent $i$ follows a similar procedure, but this time the sorting criterion is the share of agent $i$ on the interference at receiver $k \in O^{(t)}_{i}$, i.e., ${g^{(t-1)}_{i\rightarrow k}p^{(t-1)}_i}\left(\sum_{j \in N, j \neq k}g^{(t-1)}_{j\rightarrow k}p^{(t-1)}_j + \sigma^2\right)^{-1}$,
	in order to give priority to the most significantly affected interfered neighbors by agent $i$'s interference.
	
	The way we organize the local information to build $s^{(t)}_i$ accommodates some intuitive and systematic basics. Based on these basics, we perfected our design by trial-and-error with some preliminary simulations. We now describe the state of agent $i$ at time slot $t$, i.e., $s^{(t)}_i$, by dividing it into three main feature groups as:
	\subsubsection{Local Information} The first element of this feature group is agent $i$'s transmit power during previous time slot, i.e., $p^{(t-1)}_i$. Then, this is followed by the second and third elements that specify agent $i$'s most recent potential contribution on the network objective \eqref{eq:DynOptProblem}: ${1}/{w^{(t)}_i}$ and $C^{(t-1)}_i$. For the second element, we do not directly use $w^{(t)}_i$ which tends to be quite large as $\bar{C}^{(t)}_i$ is close to zero from \eqref{eq:pfsweight}. We found that using ${1}/{w^{(t)}_i}$ is more desirable. Finally, the last four elements of this feature group are the last two measurements of its direct downlink channel and the total interference-plus-noise power at receiver $i$: $g^{(t)}_{i\rightarrow i}$, $g^{(t-1)}_{i\rightarrow i}$, $\sum_{j \in N, j \neq i}g^{(t)}_{j\rightarrow i}p^{(t-1)}_j + \sigma^2$, and $\sum_{j \in N, j \neq i}g^{(t-1)}_{j\rightarrow i}p^{(t-2)}_j + \sigma^2$. Hence, a total of seven input ports of the input layer are reserved for this feature group. In our state set design, we take the last two measurements into account to give the agent a better chance to track its environment change. Intuitively, the lower the maximum Doppler frequency, the slower the environment changes, so that having more past measurements will help the agent to make better decisions \cite{yu2017deep}. On the other hand, this will result with having more state information which may increase the complexity and decrease the learning efficiency. Based on preliminary simulations, we include two past measurements. 
	\subsubsection{Interfering Neighbors} This feature group lets agent $i$ observe the interference from its neighbors to receiver $i$ and what is the contribution of these interferers on the objective \eqref{eq:DynOptProblem}. For each interferer $j \in \bar{I}^{(t)}_{i}$, three input ports are reserved for $g^{(t)}_{j\rightarrow i}p^{(t-1)}_j$, ${1}/{w^{(t-1)}_j}$, $C^{(t-1)}_j$. The first term indicates the interference that agent $i$ faced from its interferer $j$; the other two terms imply the significance of agent $j$ in the objective \eqref{eq:DynOptProblem}. Similar to the local information feature explained in the previous paragraph, agent $i$ also considers the history of its interferers in order to track changes in its own receiver's interference condition. For each interferer $j' \in \bar{I}^{(t-1)}_{i}$, three input ports are reserved for $g^{(t-1)}_{j'\rightarrow i}p^{(t-2)}_{j'}$, ${1}/{w^{(t-2)}_{j'}}$, $C^{(t-2)}_{j'}$. A total of $6c$ state elements are reserved for this feature group.
	\subsubsection{Interfered Neighbors} Finally, agent $i$ uses the feedback from its interfered neighbors to gauge its interference to nearby receivers and the contribution of them on the objective \eqref{eq:DynOptProblem}. If agent $i$'s link was inactive during the previous time slot, then $O^{(t-1)}_{i} = \emptyset$. For this case, if we ignore the history and directly consider the current interfered neighbor set, the corresponding state elements will be useless. Note that agent $i$'s link became inactive when its own estimated contribution on the objective \eqref{eq:DynOptProblem} was not significant enough compared to its interference to its interfered neighbors. Thus, after agent $i$'s link became inactive, in order to decide when to reactivate its link, it should keep track of the interfered neighbors that implicitly silenced itself. We solve this issue by defining time slot $t'_i$ which is the last time slot agent $i$ was active. The agent $i$ carries the feedback from interfered $k \in \bar{O}^{(t'_i)}_{i}$. We also pay attention to the fact that if $t'_i<t-1$, interfered $k$ has no knowledge of $g^{(t-1)}_{i\rightarrow k}$, but it is still able to send its local information to agent $i$. Therefore, agent $i$ reserves four elements of its state set for each interfered $k\in O^{(t'_i)}_{i}$ as $g^{(t-1)}_{k\rightarrow k}$, ${1}/{w^{(t-1)}_k}$, $C^{(t-1)}_k$, and 
	${g^{(t'_i)}_{i\rightarrow k}p^{(t'_i)}_i}\left(\sum_{j \in N, j \neq k}g^{(t-1)}_{j\rightarrow k}p^{(t-1)}_j + \sigma^2\right)^{-1}$. This makes a total of $4c$ elements of the state set reserved for the interfered neighbors.	
	\subsection{Actions}\label{sec:actions}
	Unlike taking discrete steps on the previous transmit power level (see, e.g., \cite{ghadimi2017dynamicpower}), we use discrete power levels taken between $0$ and $P_{\textrm{max}}$. All agents have the same action space, i.e., $A_i = A_j = A$, $\forall i,j \in N$. Suppose we have $|A|>1$ discrete power levels. Then, the action set is given by
	\begin{align}\label{eq:action}
	\begin{split}
	A &= \left\{0, \frac{P_{\textrm{max}}}{|A|-1}, \frac{2P_{\textrm{max}}}{|A|-1}, \dots, P_{\textrm{max}}\right\}
	\end{split}.
	\end{align}
	
	The total number of DQN output ports denoted as $N_4$ in Fig. \ref{fig:dqnmodela} is equal to $|A|$. Agent $i$ is only allowed to pick an action $a_i(t) \in A$ to update its power strategy at time slot $t$. This way of approaching the problem could increase the number of DQN output ports compared to \cite{ghadimi2017dynamicpower}, but it will increase the robustness of the learning algorithm. For example, as the maximum Doppler frequency $f_d$ or time slot duration $T$ increases, the correlation term $\rho$ in \eqref{eq:JakesModel} is going to decrease and the channel state will vary more. This situation may require the agents to react faster, i.e., possible transition from zero-power to full-power, which can be addressed efficiently with an action set composed of discrete power levels. 
	
	\subsection{Reward Function}\label{sec:reward}
	The reward function is designed to optimize the network objective \eqref{eq:DynOptProblem}. We interpret the reward as how the action of agent $i$ through time slot $t$, i.e., $p^{(t)}_i$, affects the weighted sum-rate of its own and its future interfered neighbors $O^{(t+1)}_{i}$. During the time slot $t+1$, for all agent $i \in N$, the network trainer calculates the spectral efficiency of each link $k \in O^{(t+1)}_{i}$ without the interference from transmitter $i$ as
	\begin{align}\label{eq:DynRate2}
	\begin{split}
	C^{(t)}_{k \setminus i} &= \log\left(1+\frac{g^{(t)}_{k\rightarrow k}p^{(t)}_k}{\sum_{j \neq i,k}g^{(t)}_{j\rightarrow k}p^{(t)}_j+\sigma^2}\right).
	\end{split}
	\end{align}
	The network trainer computes the term $\sum_{j \neq i,k}g^{(t)}_{j\rightarrow k}p^{(t)}_j+\sigma^2$ in \eqref{eq:DynRate2} by simply subtracting $g^{(t)}_{i\rightarrow k}p^{(t)}_i$ from the total interference-plus-noise power at receiver $k$ in time slot $t$. As assumed in Section \ref{sec:practical}, since transmitter $i\in I^{(t+1)}_k$, its interference to link $k$ in slot $t$, i.e., $g^{(t)}_{i\rightarrow k}p^{(t)}_i > \eta \sigma^2$, is accurately measurable by receiver $k$ and has been delivered to the network trainer. 
	
	In time slot $t$, we account for the externality that link $i$ causes to link $k$ using a price charged to link $i$ for generating interference to link $k$ \cite{huang2006distributedpower}:
	\begin{align}\label{eq:externality}
	\begin{split}
	\pi^{(t)}_{i\rightarrow k} &= w^{(t)}_k \left(C^{(t)}_{k \setminus i} - C^{(t)}_k \right).
	\end{split}
	\end{align}	
	Then, the reward function of agent $i \in N$ at time slot $t+1$ is defined as	
	\begin{align}\label{eq:reward}
	\begin{split}
	r^{(t+1)}_i &= w^{(t)}_i C^{(t)}_i - \sum_{k \in O^{(t+1)}_{k}} \pi^{(t)}_{i\rightarrow k}.
	\end{split}
	\end{align}
	The reward of agent $i$ consists of two main components: its direct contribution to the network objective \eqref{eq:DynOptProblem} and the penalty due to its interference to all interfered neighbors. Evidently, transmitting at peak power $p^{(t)}_i = P_{\textrm{max}}$ maximizes the direct contribution as well as the penalty, whereas being silent earns zero reward.	
	\section{Simulation Results}\label{sec:resutls}
	\subsection{Simulation Setup}\label{sec:simsetup}
\begin{figure}
		[t]
		\centering
		\subfloat[Single-link per cell with $R$ = 500 m and $r$ = 200 m.]{
			\includegraphics[width=1.0\columnwidth]{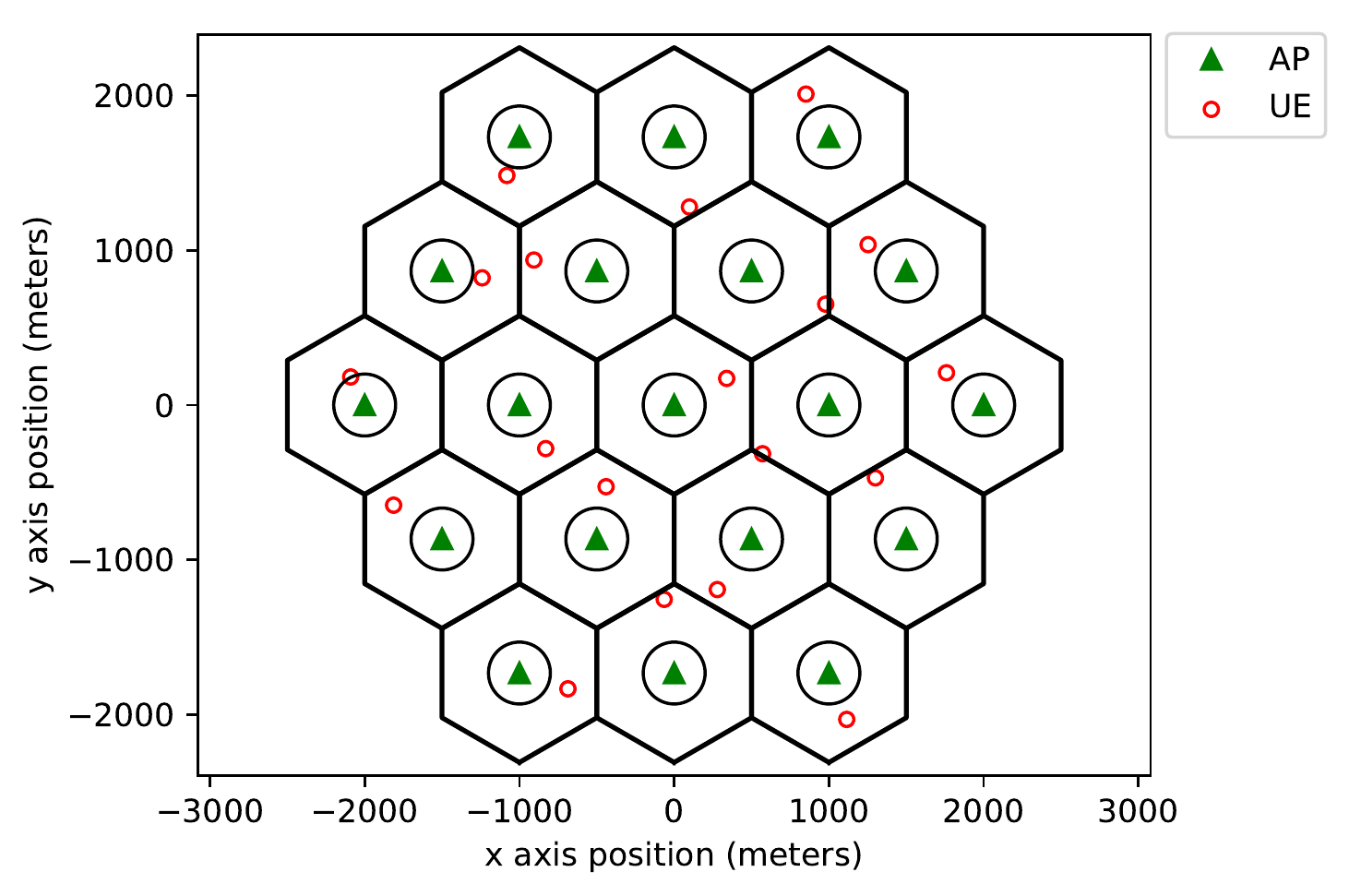}
			\label{fig:networkconfigurationsa}}
		\hfil
		\subfloat[Multi-link per cell with $R$ = 500 m and $r$ = 10 m. Each cell has a random number of links from 1 to 4 links per cell.]{
			\includegraphics[width=1.0\columnwidth]{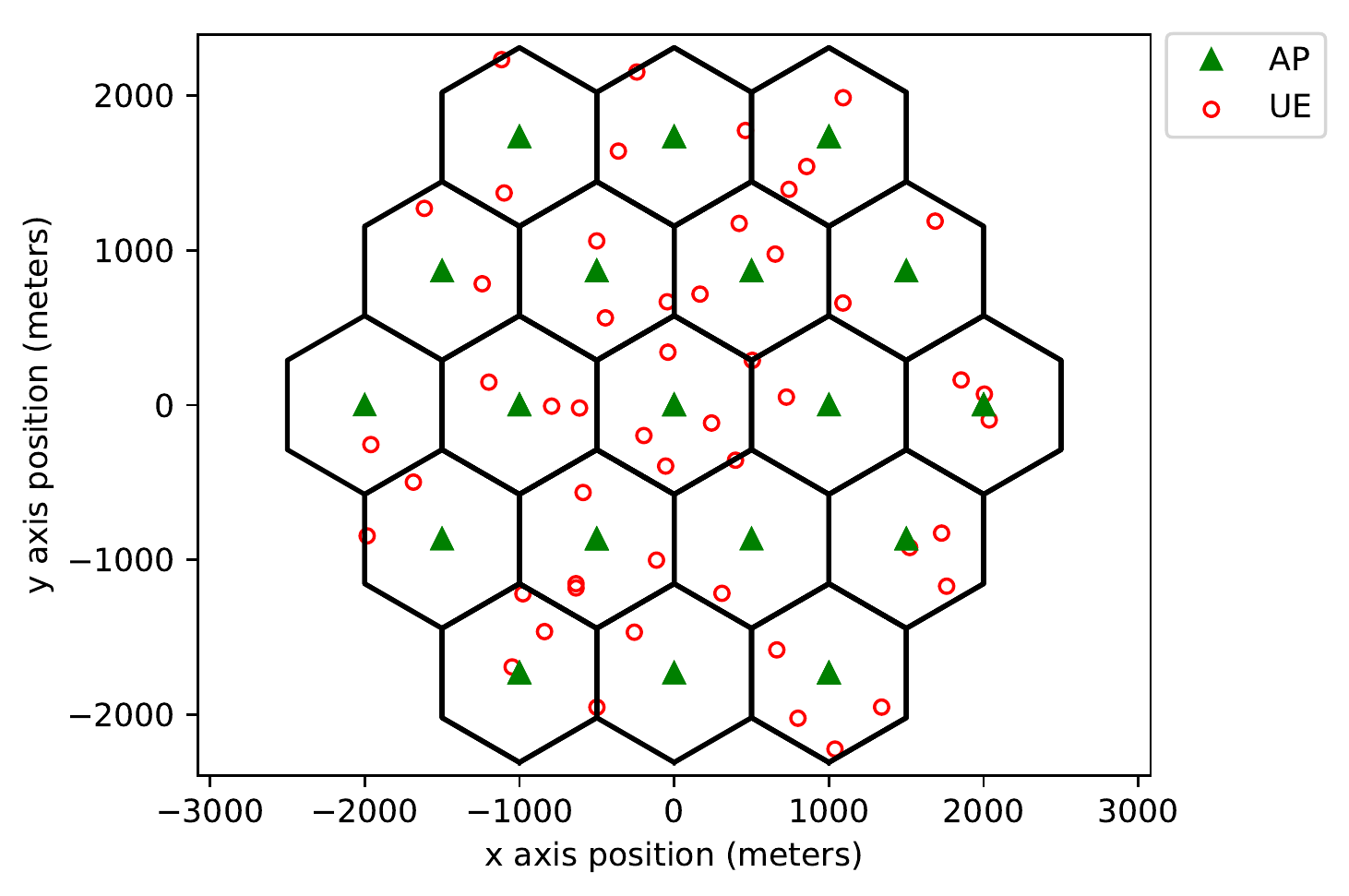}
			\label{fig:networkconfigurationsb}}
		\caption{Network configuration examples with 19 cells}
		\label{fig:networkconfigurations}
	\end{figure}

	To begin with, we consider $n$ links on $n$ homogeneously deployed cells, where we choose $n$ to be between 19 and 100. Transmitter $i$ is located at the center of cell $i$ and receiver $i$ is located randomly within the cell. We also discuss the extendability of our algorithm to multi-link per cell scenarios in Section \ref{sec:sumratemax}. The half transmitter-to-transmitter distance is denoted as $R$ and it is between 100 and 1000 meters. We also define an inner region of radius $r$ where no receiver is allowed to be placed. We set the $r$ to be between $10$ and $R-1$ meters. Receiver $i$ is placed randomly according to a uniform distribution on the area between out of the inner region of radius $r$ and the cell boundary. Fig. \ref{fig:networkconfigurations} shows two network configuration examples. 
	We set $P_{\textrm{max}}$, i.e., the maximum transmit power level of transmitter $i$, to 38 dBm over 10 MHz frequency band which is fully reusable across all links. The distance dependent path loss between all transmitters and receivers is simulated by $120.9 + 37.6\log_{10}(d)$ (in dB), where $d$ is transmitter-to-receiver distance in km. This path loss model is compliant with the LTE standard \cite{LTE-A}. The log-normal shadowing standard deviation is taken as 8 dB. The AWGN power $\sigma^2$ is -114 dBm. We set the threshold $\eta$ in \eqref{eq:InNeigh} and \eqref{eq:OutNeigh} to 5. We assume full-buffer traffic model. Similar to \cite{zhuang2016energy}, if the received SINR is greater than 30 dB, it is capped at 30 dB in the calculation of spectral efficiency by \eqref{eq:DynRate}. This is to account for typical limitations of finite-precision digital processing. In addition to these parameters, we take the period of the time-slotted system $T$ to be 20 ms. Unless otherwise stated, the maximum Doppler frequency $f_d$ is 10 Hz and identical for all receivers.
	
	We next describe the hyper-parameters used for the architecture of our algorithm. Since our goal is to ensure that the agents make their decisions as quickly as possible, we do not over-parameterize the network architecture and we use a relatively small network for training purposes. 
	Our algorithm trains a DQN with one input layer, three hidden layers, and one output layer. The hidden layers have $N_1=200$, $N_2=100$, and $N_3=40$ neurons, respectively. We have $7$ DQN input ports reserved for the local information feature group explained in Section \ref{sec:states}. The cardinality constraint on the neighbor sets $c$ is 5 agents. Hence, again from Section \ref{sec:states}, the input ports reserved for the interferer and the interfered neighbors are $6c=30$ and $4c=20$, respectively. This makes a total of $N_0=57$ input ports reserved for the state set. (We also normalize the inputs with some constants depending on $P_{\textrm{max}}$, maximum intra-cell path loss, etc., to optimize the performance.) We use ten discrete power levels, $N_4=|A|=10$. Thus, the DQN has ten outputs. Initial parameters of the DQN are generated with the truncated normal distribution function of the TensorFlow \cite{abadi2015tensorflow}. For our application, we observed that the rectifier linear unit (ReLU) function converges to a desirable power allocation slightly slower than the hyperbolic tangent (tanh) function, so we used tanh as DQN's activation function. Memory parameters at the network trainer, $M_b$ and $M_m$, are 256 and 1000 samples, respectively. We use the RMSProp algorithm \cite{ruder2016overview} with an adaptive learning rate $\alpha^{(t)}$. For a more stable deep Q-learning outcome, the learning rate is reduced as $\alpha^{(t+1)}=\lambda\alpha^{(t)}$, where $\lambda \in (0,1)$ is the decay rate of $\alpha^{(t)}$ \cite{lavet2015discountfactor}. Here, $\alpha^{(0)}$ is $5\times10^{-3}$ and $\lambda$ is $10^{-4}$. We also apply adaptive $\epsilon$-greedy algorithm: $\epsilon^{(0)}$ is initialized to 0.2 and it follows $\epsilon^{(t+1)}=\max\left\{\epsilon_{\textrm{min}},\lambda_{\epsilon}\epsilon^{(t)}\right\}$, where $\epsilon_{\textrm{min}}=10^{-2}$ and $\lambda_{\epsilon}=10^{-4}$.
	
	Although the discount factor $\gamma$ is nearly arbitrarily chosen to be close to 1 and increasing $\gamma$ potentially improves the outcomes of deep Q-learning for most of its applications \cite{lavet2015discountfactor}, we set $\gamma$ to 0.5. The reason we use a moderate level of $\gamma$ is that the correlation between agent's actions and its future rewards tends to be smaller for our application due to fading. An agent's action has impact on its own future reward through its impact on the interference condition of its neighbors and consequences of their unpredictable actions. Thus, we set $\gamma \geq 0.5$. We observed that higher $\gamma$ is not desirable either. It slows the DQN's reaction to channel changes, i.e., high $f_d$ case. For high $\gamma$, the DQN converges to a strategy that makes the links with better steady-state channel condition greedy. As $f_d$ becomes large, due to fading, the links with poor steady-state channel condition may become more advantageous for some time-slots. Having a moderate level of $\gamma$ helps detect these cases and allows poor links to be activated during these time slots when they can contribute the network objective \eqref{eq:DynOptProblem}. 
	Further, the training cycle duration $T_{u}$ is 100 time slots. After we set the parameters in \eqref{eq:DQNparameters}, we can compute the total number of DQN parameters, i.e., $|\theta|$, as 36,150 parameters. 
	After each $T_{u}$ time slots, trained parameters at the central controller will be delivered to all agents in $T_d$ time slots via backhaul network as explained in Section \ref{sec:proposedalg}. We assume that the parameters are transferred without any compression and the backhaul network uses pure peer-to-peer architecture. As $T_d = 50$ time slots, i.e., 1 second, the minimum required downlink/uplink capacity for all backhaul links is about 1 Mbps. Once the training stage is completed, the backhaul links will be used only for limited information exchange between neighbors which requires negligible backhaul link capacity. 
	
	We empirically validate the functionality of our algorithm. We implemented the proposed algorithm with TensorFlow \cite{abadi2015tensorflow}. Each result is an average of at least 10 randomly initialized simulations. We have two main phases for the simulations: training and testing. Each training lasts 40,000 time slots or $40,000\times20$ ms = 800 seconds, and each testing lasts 5,000 time slots or 100 seconds. During the testing, the trainer leaves the network and the $\epsilon$-greedy algorithm is terminated, i.e., agents stop exploring the environment.
	
	We have five benchmarks to evaluate the performance of our algorithm. The first two benchmarks are `ideal WMMSE' and `ideal FP' with instantaneous full CSI and centralized algorithm outcome. The third benchmark is the `central power allocation' (central), where we introduce one time slot delay on the full CSI and feed it to the FP algorithm. Even the single time slot delay to acquire the full CSI is a generous assumption, but it is a useful approach to reflect potential performance of negligible computation time achieved with the supervised learning approach introduced in \cite{sun2017learning}. The next benchmark is the `random' allocation, where each agent chooses its transmit power for each slot at random uniformly between 0 and $P_{\textrm{max}}$. The last benchmark is the `full-power' allocation, i.e., each agent's transmit power is $P_{\textrm{max}}$ for all slots.
	\subsection{Sum-Rate Maximization}\label{sec:sumratemax}
	In this subsection, we focus on the sum-rate by setting the weights of all network agents to 1 through all time slots.
	\begin{figure}
		[t]
		\centering
		\subfloat[Training - Moving average spectral efficiency per link of previous 250 slots.]{
			\includegraphics[width=1.0\columnwidth]{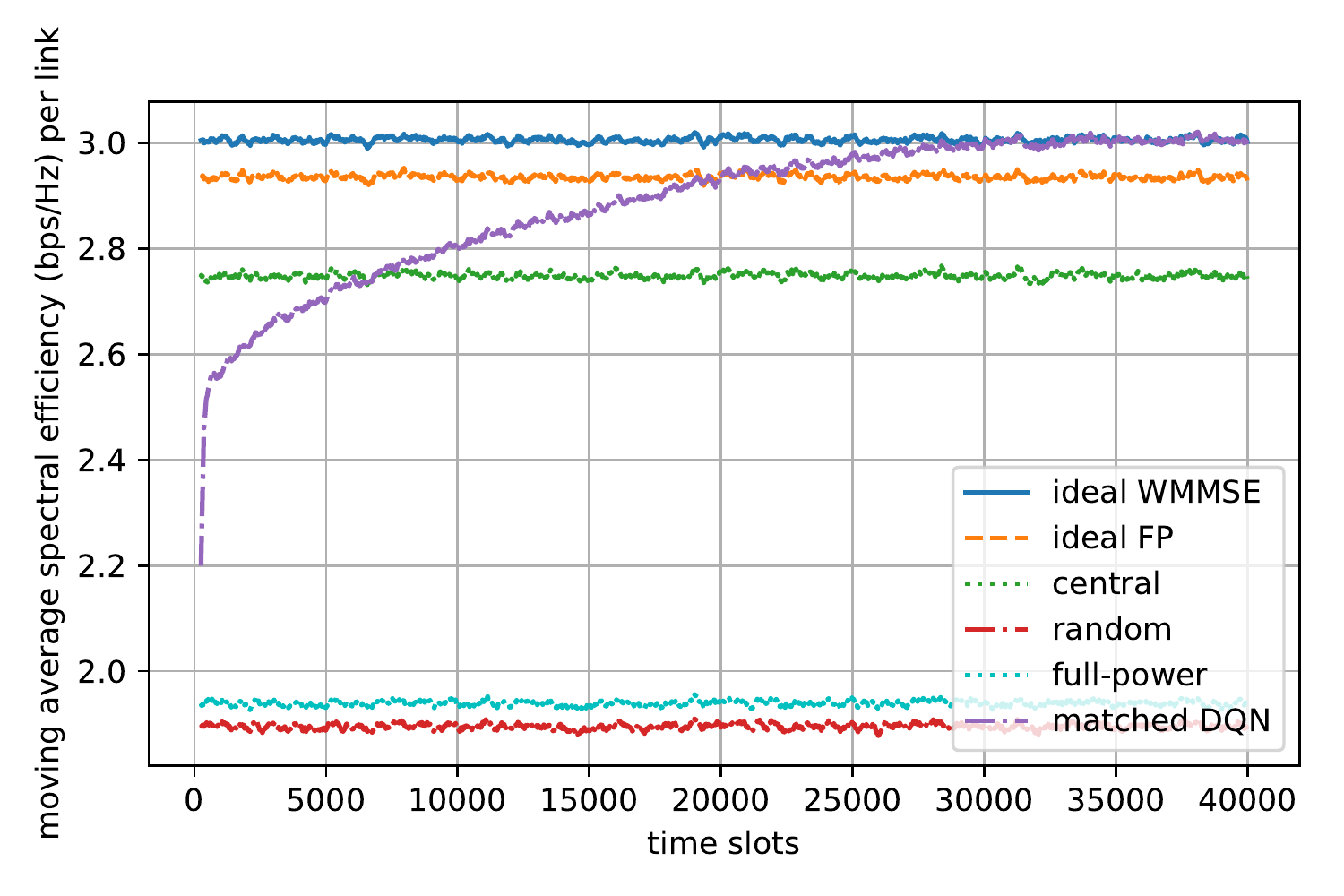}
			\label{fig:trainn19R100r10}}
		\hfil
		\subfloat[Testing - Empirical CDF.]{
			\includegraphics[width=1.0\columnwidth]{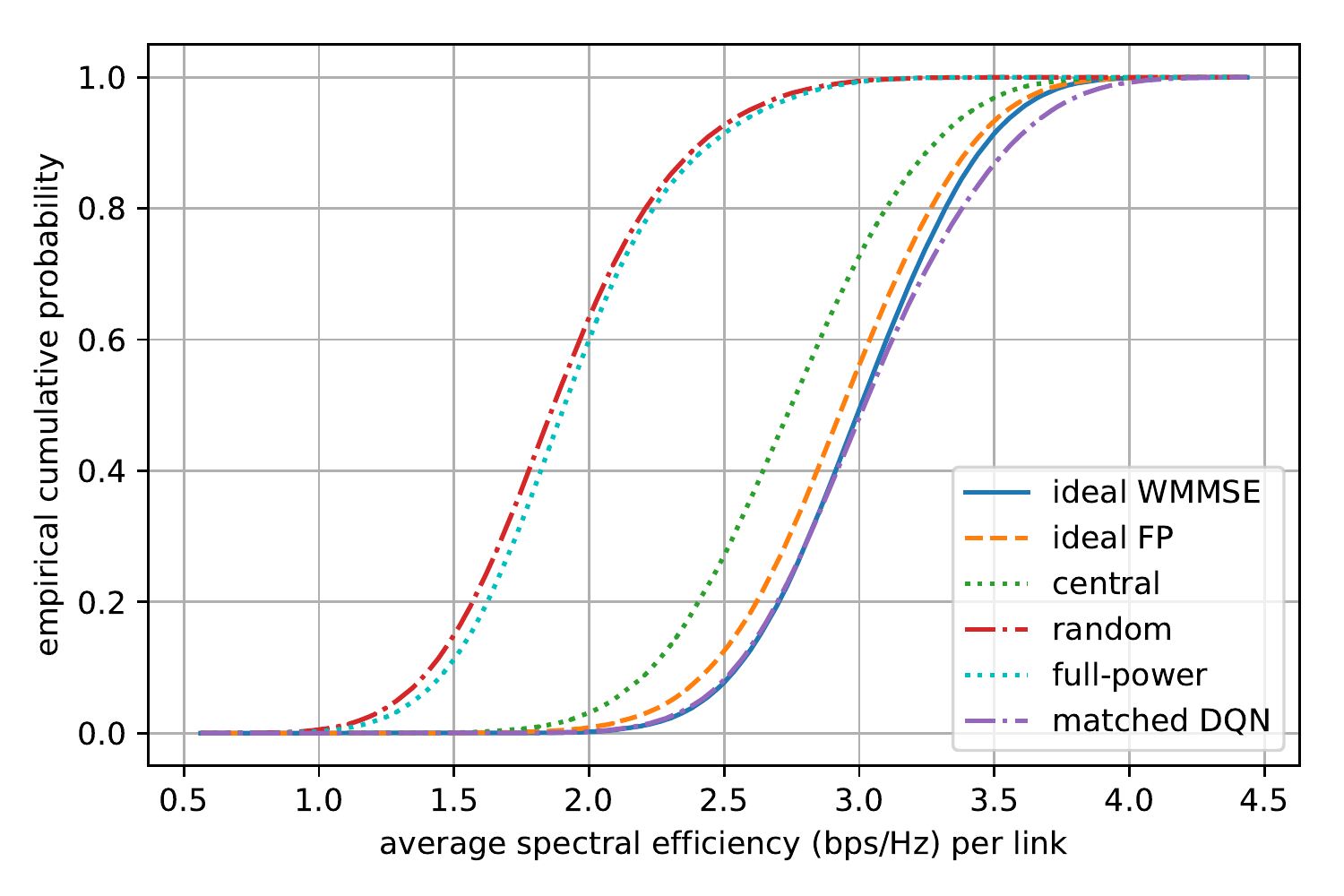}
			\label{fig:testn19R100r10}}
		\caption{Sum-rate maximization. $n$ = 19 links, $R$ = 100 m, $r$ = 10 m, $f_d$ = 10 Hz.}
		\label{fig:n19R1000r10}
	\end{figure}
	\subsubsection{Robustness}\label{sec:effectofmultiplier}	
	We fix $n=19$ links and use two approaches to evaluate performance. The first approach is the `matched' DQN where we use the first 40,000 time slots to train a DQN from scratch, whereas for the `unmatched' DQN we ignore the matched DQN specialized for a given specific initialization, and for the testing (the last 5,000 time slots) we randomly pick another DQN trained for another initialization with the same $R$ and $r$ parameters. In other words, for the unmatched DQN case, we skip the training stage and use the matched DQN that was trained for a different network initialization scenario and was stored in the memory. Here an unmatched DQN is always trained for a random initialization with $n$ = 19 links and $f_d$ = 10 Hz. 
	\begin{table}[t]
		
		\tabcolsep 0pt \caption{Testing results for variant half transmitter-to-transmitter distance. $n$ = 19 links, $r$ = 10 m, $f_d$ = 10 Hz. }
		\begin{center}
			\def\temptablewidth{1\columnwidth}
			{\rule{\temptablewidth}{1pt}}
			\begin{tabular*}{\temptablewidth}{@{\extracolsep{\fill}}|c|cc|ccccc|}
				{}&\multicolumn{7}{c|}{average sum-rate performance in bps/Hz per link}\\
				{} &  \multicolumn{2}{c|}{DQN} & \multicolumn{5}{c|}{benchmark power allocations} \\
				{$R$ (m)} &matched &unmatched &WMMSE & FP & central & random & full-power
				\\\hline \hline
				100 & 3.04 & 2.83 & 3.01 & 2.94 & 2.75 & 1.89 & 1.94 \\
				300 & 2.76 & 2.49 & 2.69 & 2.61 & 2.46 & 1.45 & 1.47 \\
				400 & 2.80 & 2.49 & 2.70 & 2.63 & 2.48 & 1.40 & 1.42 \\
				500 & 2.78 & 2.50 & 2.66 & 2.58 & 2.44 & 1.36 & 1.37 \\
				1000 & 2.71 & 2.43 & 2.61 & 2.54 & 2.40 & 1.31 & 1.33 
			\end{tabular*}
			{\rule{\temptablewidth}{1pt}}
		\end{center}
		\label{table:robustness1}
	\end{table}
	\begin{table}[t]	
		\tabcolsep 0pt \caption{Testing results for variant inner region radius. $n$ = 19 links, $R$ = 500 m, $f_d$ = 10 Hz. }
		\begin{center}
			\def\temptablewidth{1\columnwidth}
			{\rule{\temptablewidth}{1pt}}
			\begin{tabular*}{\temptablewidth}{@{\extracolsep{\fill}}|c|cc|ccccc|}
				{}&\multicolumn{7}{c|}{average sum-rate performance in bps/Hz per link}\\
				{} &  \multicolumn{2}{c|}{DQN} & \multicolumn{5}{c|}{benchmark power allocations} \\
				{$r$ (m)} &matched &unmatched &WMMSE & FP & central & random & full-power
				\\\hline \hline
				10 & 2.78 & 2.50 & 2.66 & 2.58 & 2.44 & 1.36 & 1.37 \\
				200 & 2.33 & 2.04 & 2.28 & 2.20 & 2.06 & 0.92 & 0.93 \\
				400 & 2.06 & 1.84 & 2.00 & 1.93 & 1.80 & 0.70 & 0.70 \\
				499 & 2.09 & 1.87 & 2.05 & 1.98 & 1.84 & 0.65 & 0.64
			\end{tabular*}
			{\rule{\temptablewidth}{1pt}}
		\end{center}
		\label{table:robustnessr}
	\end{table}
	\begin{table}[t]	
		\tabcolsep 0pt \caption{Testing results for variant maximum Doppler frequency. $n$ = 19 links, $R$ = 500 m, $r$ = 10 m. (`random' means $f_d$ of each link is randomly picked between 2 Hz and 15 Hz for each time slot $t$. `uncorrelated' means that we set $f_d \to \infty$ and $\rho$ becomes zero.) }
		\begin{center}
			\def\temptablewidth{1\columnwidth}
			{\rule{\temptablewidth}{1pt}}
			\begin{tabular*}{\temptablewidth}{@{\extracolsep{\fill}}|c|cc|ccccc|}
				{}&\multicolumn{7}{c|}{average sum-rate performance in bps/Hz per link}\\
				{} &  \multicolumn{2}{c|}{DQN} & \multicolumn{5}{c|}{benchmark power allocations} \\
				{$f_d$ (Hz)} &matched &unmatched &WMMSE & FP & central & random & full-power
				\\\hline \hline
				2 & 2.80 & 2.48 & 2.64 & 2.55 & 2.54 & 1.36 & 1.37 \\
				5 & 2.83 & 2.47 & 2.68 & 2.58 & 2.52 & 1.21 & 1.21 \\
				10 & 2.78 & 2.50 & 2.66 & 2.58 & 2.44 & 1.36 & 1.37 \\
				15 & 2.85 & 2.45 & 2.72 & 2.64 & 2.47 & 1.35 & 1.36 \\
				random & 2.88 & 2.55 & 2.80 & 2.71 & 2.59 & 1.47 & 1.49 \\
				uncorrelated 
				& 2.82 & 2.41 & 2.68 & 2.61 & 2.39 & 1.55 & 1.57
			\end{tabular*}
			{\rule{\temptablewidth}{1pt}}
		\end{center}
		\label{table:robustnessfd}
	\end{table}
	
	In Table \ref{table:robustness1}, we vary $R$ and see that training a DQN from scratch for the specific initialization is able to outperform both state-of-the-art centralized algorithms that are under ideal conditions such as full CSI and no delay. Interestingly, the unmatched DQN approach converges to the central power allocation where we feed the FP algorithm with delayed full CSI. The DQN approach achieves this performance with distributed execution and incomplete CSI. In addition, training a DQN from scratch enables our algorithm to learn to compensate for CSI delays and specialize for its network initialization scenario. Training a DQN from scratch swiftly converges in about 25,000 time slots (shown in Fig. \ref{fig:trainn19R100r10}). 
	
	Additional simulations with $r$ and $f_d$ taken as variables are summarized in Table \ref{table:robustnessr} and Table \ref{table:robustnessfd}, respectively. As the area of receiver-free inner region increases, the receivers get closer to the interfering transmitters and the interference mitigation becomes more necessary. Hence, the random and full-power allocations tend to show much lower sum-rate performance compared to the central algorithms. For that case, our algorithm still shows decent performance and the convergence rate is still about 25,000 time slots. We also stress the DQN under various $f_d$ scenarios. As we reduce $f_d$, its sum-rate performance remains unchanged, but the convergence time drops to 15,000 time slots. As $f_d \to \infty$, i.e., we set $\rho = 0$ to remove the temporal correlation between current channel condition and past channel conditions, the convergence takes more than 35,000 time slots. Intuitively, the reason of this effect on the convergence rate is that the variation of states visited during the training phase is proportional to $f_d$. Further, the comparable performance of the unmatched DQN with the central power allocation shows the robustness of our algorithm to the changes in interference conditions and fading characteristics of the environment.
	\subsubsection{Scalability}\label{sec:scalability}
	In this subsection, we increase the total number of links to investigate the scalability of our algorithm. 
	As we increase $n$ to 50 links, the DQN still converges in 25,000 time slots with high sum-rate performance. As we keep on increasing $n$ to 100 links, from Table \ref{table:scalability}, the matched DQN's sum-rate outperformance drops because of the fixed input architecture of the DQN. 
	\begin{table}[t]	
		\tabcolsep 0pt \caption{Testing results for variant total number of links. $R$ = 500 m, $r$ = 10 m, $f_d$ = 10 Hz.}
		\begin{center}
			\def\temptablewidth{1\columnwidth}
			{\rule{\temptablewidth}{1pt}}
			\begin{tabular*}{\temptablewidth}{@{\extracolsep{\fill}}|c|cc|ccccc|}
				{}&\multicolumn{7}{c|}{average sum-rate performance in bps/Hz per link}\\
				{} &  \multicolumn{2}{c|}{DQN} & \multicolumn{5}{c|}{benchmark power allocations} \\
				{$n$ (links)} &matched &unmatched &WMMSE & FP & central & random & full-power
				\\\hline \hline
				19 & 2.78 & 2.50 & 2.66 & 2.58 & 2.44 & 1.36 & 1.37 \\
				50 & 2.28 & 1.99 & 2.17 & 2.13 & 2.00 & 1.01 & 1.02 \\
				100 & 1.92 & 1.68 & 1.90 & 1.88 & 1.74 & 0.87 & 0.89
			\end{tabular*}
			{\rule{\temptablewidth}{1pt}}
		\end{center}
		\label{table:scalability}
	\end{table}
	Note that each agent only considers $c = 5$ interferer and interfered neighbors. The performance of DQN can be improved for that case by increasing $c$ at a higher computational complexity. Additionally, the unmatched DQN trained for just 19 links still shows good performance as we increase the number of links. 
	
	It is worth pointing out that each agent is able to determine its own action in less than $0.5$ ms on a personal computer. Therefore, our algorithm is suitable for dynamic power allocation. In addition, running a single batch takes less than $T$ = 20 ms. Most importantly, because of the fixed architecture of the DQN, increasing the total number of links from 19 to 100 has no impact on these values. It will just increase the queue memory in the network trainer. For the FP algorithm it takes about 15 ms to converge for $n$ = 19 links, but with $n$ = 100 links it becomes 35 ms. The WMMSE algorithm converges slightly slower, and the convergence time is still proportional to $n$ which limits its scalability.	
	\begin{figure}
		[t]
		\centering
		\subfloat[Training - Moving average spectral efficiency per link of previous 250 slots.]{
			\includegraphics[width=1.0\columnwidth]{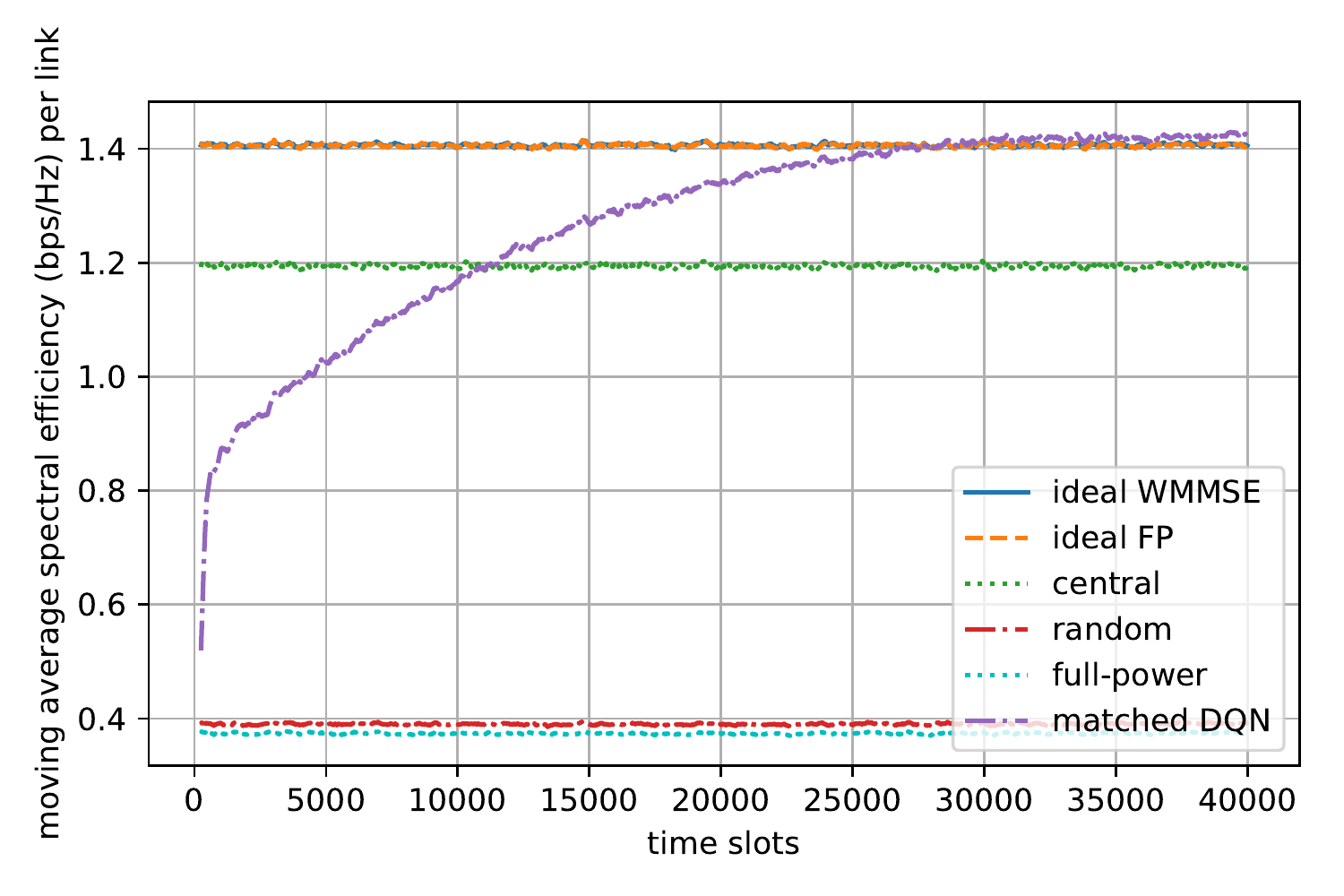}
			\label{fig:trainn19R500r10ueperbs2}}
		\hfil
		\subfloat[Testing - Empirical CDF.]{
			\includegraphics[width=1.0\columnwidth]{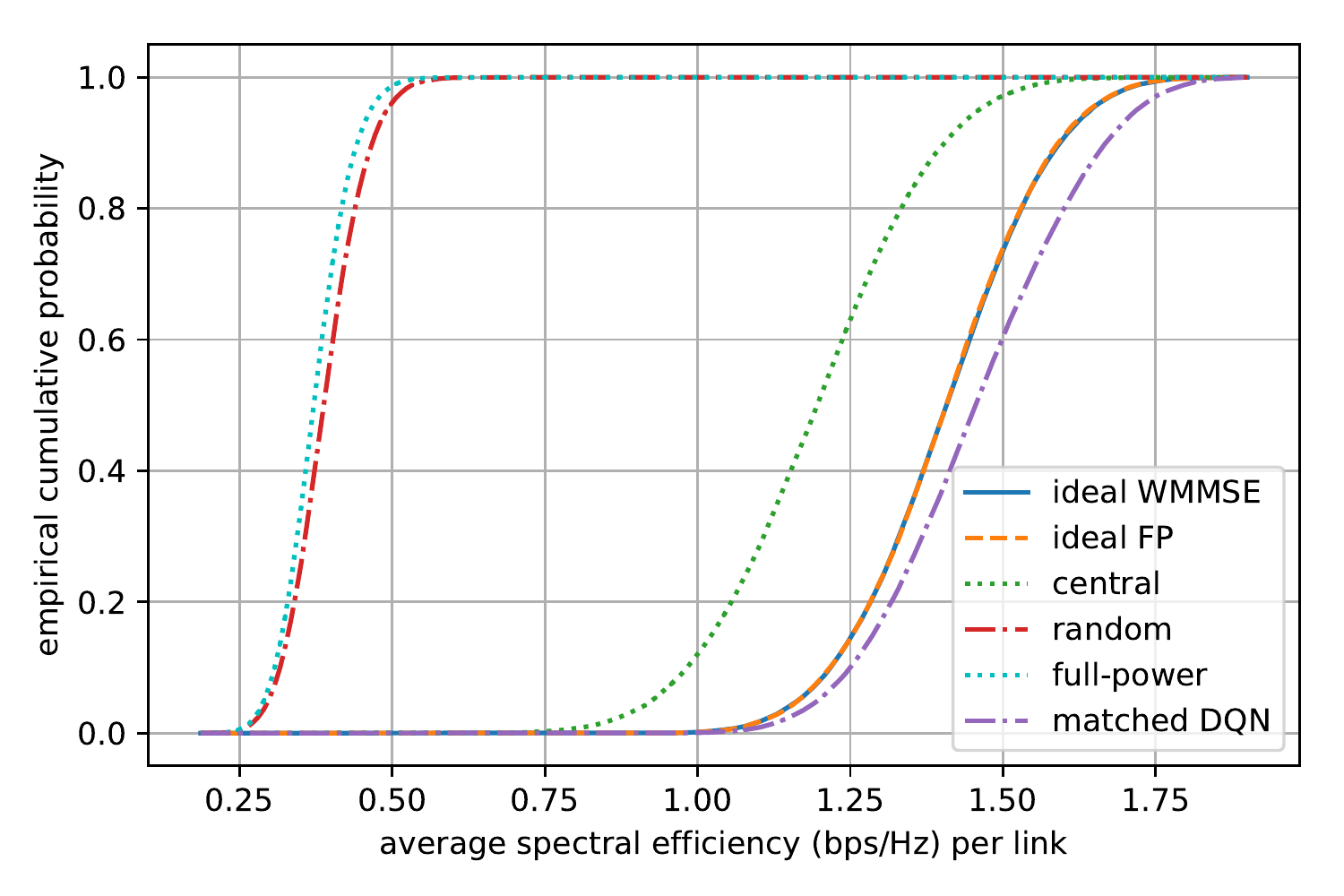}
			\label{fig:testn19R500r10ueperbs2}}
		\caption{Sum-rate maximization. 4 links per cell scenario. UMi street canyon. $n$ = 76 links deployed on 19 cells, $R$ = 500 m, $r$ = 10 m, $f_d$ = 10 Hz.}
		\label{fig:n19R500r10ueperbs2}
	\end{figure}
	\begin{table}[t]	
		\tabcolsep 0pt \caption{Testing results for variant number of links per cell. 19 cells, $R$ = 500 m, $r$ = 10$\,\,$m.}
		\begin{center}
			\def\temptablewidth{1\columnwidth}
			{\rule{\temptablewidth}{1pt}}
			\begin{tabular*}{\temptablewidth}{@{\extracolsep{\fill}}|c|cc|ccccc|}
				{}&\multicolumn{7}{c|}{average sum-rate performance in bps/Hz per link}\\
				{} &  \multicolumn{2}{c|}{DQN} & \multicolumn{5}{c|}{benchmark power allocations} \\
				{links per cell} &matched &unmatched &WMMSE & FP & central & random & full-power
				\\\hline \hline
				2 & 1.84 & 1.58 & 1.78 & 1.74 & 1.59 & 0.58 & 0.57 \\
				4 & 1.25 & 1.06 & 1.24 & 1.22 & 1.10 & 0.25 & 0.25 \\
				random & 1.61 & 1.37 & 1.57 & 1.53 & 1.40 & 0.44 & 0.44
			\end{tabular*}
			{\rule{\temptablewidth}{1pt}}
		\end{center}
		\label{table:multilink}
	\end{table}	
	\begin{table}[t]	
		
		\tabcolsep 0pt \caption{Testing results for variant number of links per cell and UMi street canyon model. 19 cells, $R$ = 500 m, $r$ = 10$\,\,$m.}
		\begin{center}
			\def\temptablewidth{1\columnwidth}
			{\rule{\temptablewidth}{1pt}}
			\begin{tabular*}{\temptablewidth}{@{\extracolsep{\fill}}|c|cc|ccccc|}
				{}&\multicolumn{7}{c|}{average sum-rate performance in bps/Hz per link}\\
				{} &  \multicolumn{2}{c|}{DQN} & \multicolumn{5}{c|}{benchmark power allocations} \\
				{links per cell} &matched &unmatched &WMMSE & FP & central & random & full-power
				\\\hline \hline
				2 & 2.60 & 2.29 & 2.53 & 2.52 & 2.27 & 1.04 & 0.99 \\
				4 & 1.46 & 1.23 & 1.41 & 1.41 & 1.19 & 0.39 & 0.37 \\
				random & 2.09 & 1.78 & 2.01 & 2.01 & 1.77 & 0.78 & 0.76
			\end{tabular*}
			{\rule{\temptablewidth}{1pt}}
		\end{center}
		\label{table:multilinkUMi}
		
	\end{table}
	\subsubsection{Extendability to Multi-Link per Cell Scenarios and Different Channel Models}\label{sec:extendability}
	In this subsection, we first consider a special homogeneous cell deployment case with co-located transmitters at the cell centers. We also assume that the co-located transmitters within a cell do not perform successive interference cancellation \cite{sun2017learning}. The WMMSE and FP algorithms can be applied to this multi-link per cell scenario without any modifications.
	
	We fix $R$ and $r$ to 500 and 10 meters, respectively. We set $f_d$ to 10 Hz and the total number of cells to $19$. We first consider two scenarios where each cell has 2 and 4 links, respectively. The third scenario assigns each cell a random number of links from 1 to 4 links per cell as shown in Fig. \ref{fig:networkconfigurationsb}. The testing stage results for these multi-link per cell scenarios are given in Table \ref{table:multilink}. As shown in Table \ref{table:multilinkUMi}, we further test these scenarios using a different channel model called urban micro-cell (UMi) street canyon model of \cite{tr38901}. For this model, we take the carrier frequency as 1 GHz. The transmitter and receiver antenna heights are assumed to be 10 and 1.5 meters, respectively.
	
	Our simulations for these scenarios show that as we increase number of links per cell, the training stage still converges in about 25,000 time slots. Fig. \ref{fig:trainn19R500r10ueperbs2} shows the convergence rate of training stage for 4 links per cell scenario with 76 links. In Fig. \ref{fig:trainn19R500r10ueperbs2}, we also show that using a different channel model, i.e., UMi street canyon, does not affect the convergence rate. Although the convergence rate is unaffected, the proposed algorithm's average sum-rate performance decreases as we increase number of links per cell. Our algorithm still outperforms the centralized algorithms even for 4 links per cell scenario for both channel models. Another interesting fact is that although the unmatched DQN was trained for a single-link deployment scenario and can not handle the delayed CSI constraint as good as the matched DQN, it gives comparable performance with the `central' case. Thus, the unmatched DQN is capable of finding good estimates of optimal actions for unseen local state inputs.
	\subsection{Proportionally Fair Scheduling}\label{sec:pfs}
	\begin{table}[t]		
		\tabcolsep 0pt \caption{Proportional fair scheduling with variant half transmitter-to-transmitter distance. $n$ = 19 links, $r$ = 10 m, $f_d$ = 10 Hz.}
		\begin{center}
			\def\temptablewidth{1\columnwidth}
			{\rule{\temptablewidth}{1pt}}
			\begin{tabular*}{\temptablewidth}{@{\extracolsep{\fill}}|c|cc|ccccc|}
				{}&\multicolumn{7}{c|}{convergence of the network sum log-average rate $\left(\ln\left(\textrm{bps}\right)\right)$}\\
				{} &  \multicolumn{2}{c|}{DQN} & \multicolumn{5}{c|}{benchmark power allocations}\\
				{$R$ (m)} &matched & unmatched & WMMSE & FP & central & random & full-power
				\\\hline \hline
				100 & 26.25 & 24.75 & 29.12 & 28.27 & 25.21 & 15.03 & 14.36 \\
				300 & 22.95 & 21.53 & 23.80 & 23.31 & 20.57 & -2.64 & -4.88 \\
				400 & 22.72 & 20.91 & 22.64 & 22.48 & 19.85 & -7.52 & -10.05 \\
				500 & 21.25 & 18.45 & 20.69 & 20.88 & 18.19 & -11.76 & -14.59 \\
				1000 & 18.37 & 14.67 & 17.27 & 17.34 & 14.53 & -16.66 & -19.64
			\end{tabular*}
			{\rule{\temptablewidth}{1pt}}
		\end{center}
		\label{table:PFSR}
	\end{table}
	\begin{table}[t]	
		\tabcolsep 0pt \caption{Proportional fair scheduling with variant inner region radius. $n$ = 19 links, $R$ = 500 m, $f_d$ = 10 Hz. }
		\begin{center}
			\def\temptablewidth{1\columnwidth}
			{\rule{\temptablewidth}{1pt}}
			\begin{tabular*}{\temptablewidth}{@{\extracolsep{\fill}}|c|cc|ccccc|}
				{}&\multicolumn{7}{c|}{convergence of the network sum log-average rate $\left(\ln\left(\textrm{bps}\right)\right)$}\\
				{} &  \multicolumn{2}{c|}{DQN} & \multicolumn{5}{c|}{benchmark power allocations} \\
				{$r$ (m)} &matched &unmatched &WMMSE & FP & central & random & full-power
				\\\hline \hline
				10  & 21.25 & 18.45 & 20.69 & 20.88 & 18.19 & -11.76 & -14.59 \\
				200 & 20.24 & 17.78 & 19.01 & 19.25 & 16.58 & -16.31 & -19.43 \\
				400 & 16.65 & 14.82 & 16.70 & 16.84 & 13.92 & -26.82 & -30.35 \\
				499 & 13.99 & 12.43 & 14.12 & 14.60 & 11.56 & -35.46 & -39.29 
			\end{tabular*}
			{\rule{\temptablewidth}{1pt}}
		\end{center}
		\label{table:PFSr}
	\end{table}
	In this subsection, we change the link weights according to \eqref{eq:pfsweight} to ensure fairness as described in Section \ref{sec:dynpowcontrol}. We choose the $\beta$ term in \eqref{eq:exponentialmovingaveragerate} to be 0.01 and use convergence to the objective in \eqref{eq:pfsobjective} as performance-metric of the DQN. 
	We also make some additions to the training and testing stage of DQN. We need an initialization for the link weights. This is done by letting all transmitters to serve their receivers with full-power at $t$ = 0, and initialize weights according to the initial spectral efficiencies computed from \eqref{eq:DynRate}. For the testing stage, we reinitialize the weights after the first 40,000 slots to see whether the trained DQN can achieve fairness as fast as the centralized algorithms.
	\begin{figure}
		[t]
		\centering
		\subfloat[Training.]{
			\includegraphics[width=1.0\columnwidth]{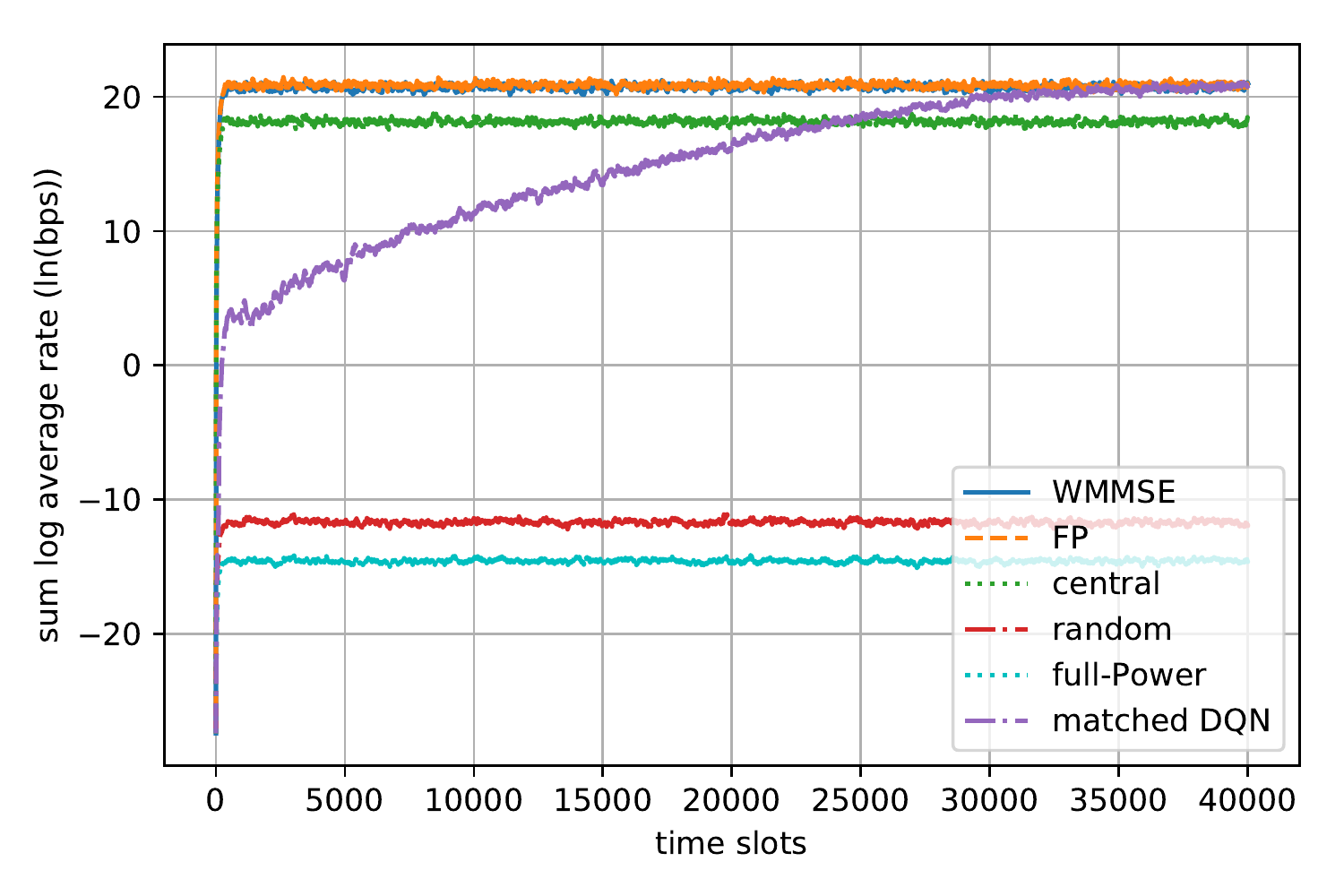}
			\label{fig:PFSTrainN19R500r10}}
		\hfil
		\subfloat[Testing.]{
			\includegraphics[width=1.0\columnwidth]{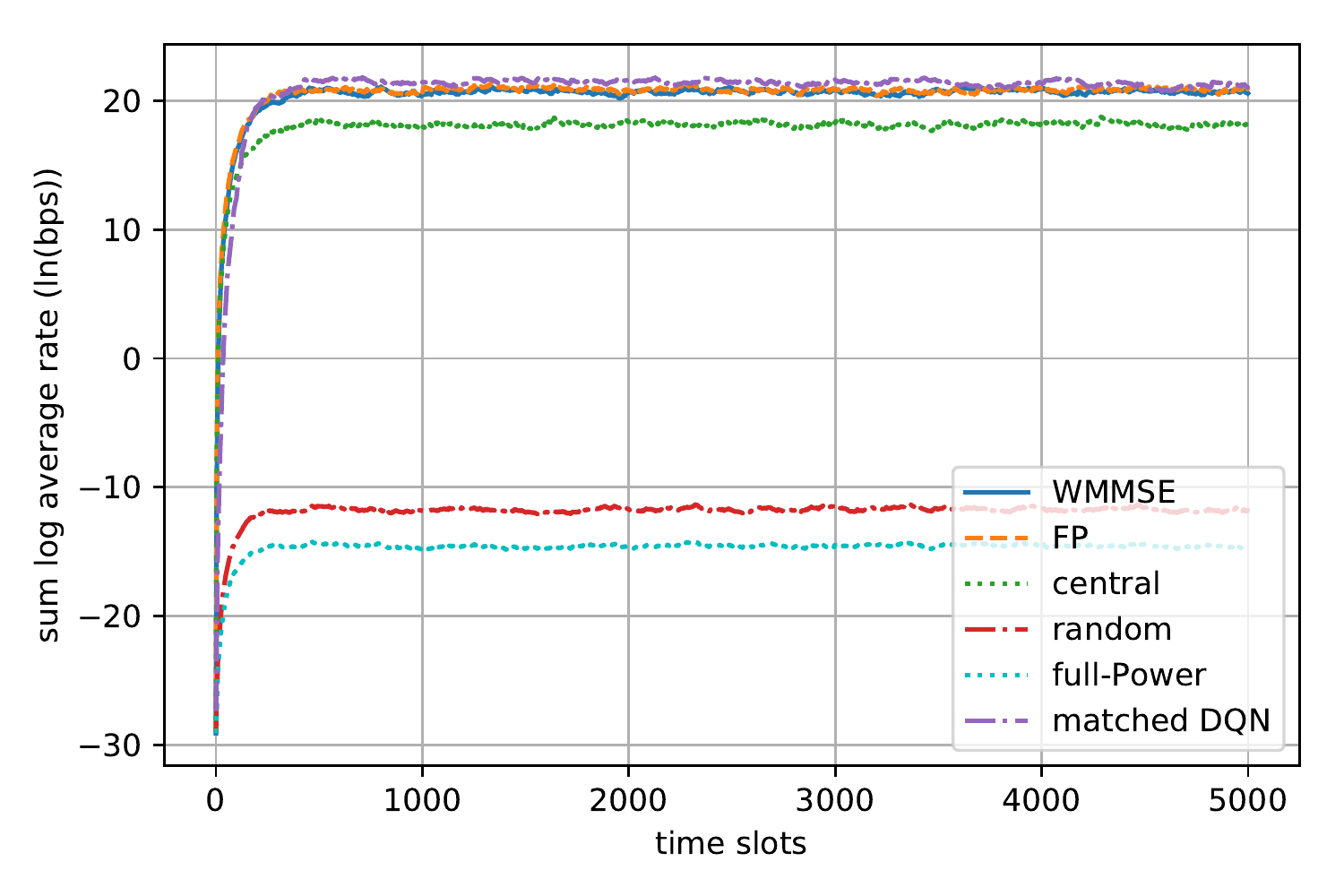}
			\label{fig:PFSTestN19R500r10}}
		\caption{Proportionally fair scheduling. $n$ = 19 links, $R$ = 500 m, $r$ = 10 m, $f_d$ = 10 Hz.}
		\label{fig:PFSN19R500r10}
	\end{figure}
	
	As shown in Fig. \ref{fig:PFSN19R500r10}, the training stage converges to a desirable scheduling in about 30,000 time slots. Once the network is trained, as we reinitialize the link weights, our algorithm converges to an optimal scheduling in a distributed fashion as fast as the centralized algorithms. Next, we set $R$ and $r$ as variables to get results in Table \ref{table:PFSR} and Table \ref{table:PFSr}. We see that the trained DQN from scratch still outperforms the centralized algorithms in most of the initializations, using the unmatched DQN also achieves a high performance similar to the previous sections.
	\section{Conclusion and Future Work}\label{sec:conclusion}
	In this paper, we have proposed a distributively executed model-free power allocation algorithm which outperforms or achieves comparable performance with existing state-of-the-art centralized algorithms. We see potentials in applying the reinforcement learning techniques on various dynamic wireless network resource management tasks in place of the optimization techniques. The proposed approach returns the new suboptimal power allocation much quicker than two of the popular centralized algorithms taken as the benchmarks in this paper. In addition, by using the limited local CSI and some realistic practical constraints, our deep Q-learning approach usually outperforms the generic WMMSE and FP algorithms which requires the full CSI which is an inapplicable condition. Differently from most advanced optimization based power control algorithms, e.g. WMMSE and FP, that require both instant and accurate measurements of individual channel gains, our algorithm only requires accurate measurements of some delayed received power values that are higher than a certain threshold above noise level. An extension to an imperfect CSI case with inaccurate CSI measurements is left for future work. 
	
	Meng \emph{et al.} \cite{meng2018deepmulti} is an extension of our preprint version \cite{nasir2018deep} to multiple users in a cell, which is also addressed in the current paper. Although the centralized training phase seems to be a limitation on the proposed algorithm in terms of scalability, we have shown that a DQN trained for a smaller wireless network can be applied to a larger wireless network and a jump-start on the training of DQN can also be implemented by using initial parameters taken from another DQN previously trained for a different setup.
	
	Finally, we used global training in this paper, whereas reinitializing a local training over the regions where new links joined or performance dropped under a certain threshold is also an interesting direction to consider. Besides the regional training, completely distributed training can be considered, too. While a centralized training approach saves computational resources and converges faster, distributed training may beat a path for an extension of the proposed algorithm to some other channel deployment scenarios that involves mobile users. The main hurdle on the way to apply distributed training is to avoid the instability caused by the environment non-stationarity. 
	\section{Acknowledgement}
	We thank Dr. Mingyi Hong, Dr. Wei Yu, Dr. Georgios Giannakis, and Dr. Gang Qian for stimulating discussions.
	
	\bibliographystyle{IEEEtran}
	\bibliography{IEEEabrv}
\end{document}